\documentclass{aa}  
\usepackage{graphicx}
\usepackage[colorlinks,allcolors=blue]{hyperref}
\bibpunct{(}{)}{;}{a}{}{,}
\usepackage{bm,natbib,url,wasysym}
\usepackage[varg]{txfonts}
\usepackage{xcolor}
\usepackage{amsmath}

\definecolor{dgreen}{rgb}{0.0, 0.6, 0.0}
                                
\newcommand{\sect}[1]{Sect.\,\ref{#1}}
\newcommand{\app}[1]{Appendix\,\ref{#1}}
\newcommand{\sects}[1]{Sects.\,\ref{#1}}
\newcommand{\fig}[1]{Fig.\,\ref{#1}}

\newcommand{\tab}[1]{Table\,\ref{#1}}

\newcommand{\eqn}[1]{Eq.\,(\ref{#1})}
\newcommand{\eqs}[1]{Eqs.\,(\ref{#1})}

\makeatletter
\renewcommand*\aa@pageof{, page \thepage{} of \pageref*{LastPage}}
\makeatother
\begin{document} 

\titlerunning{}
\authorrunning{}

   \title{Stellar X-rays and magnetic activity in 3D MHD coronal models}

   \author{J. Zhuleku,
           J. Warnecke
          \and
          H. Peter
          }

   \institute{Max Planck Institute for Solar System Research, 37077 G\"ottingen, Germany\\
              \email{zhuleku@mps.mpg.de}}

   \date{Received/Accepted}

  \abstract
  %
  %
  {Observations suggest a power-law relation between the coronal emission in X-rays, $L_{\rm{X}}$, and the total (unsigned)  magnetic flux at the stellar surface, $\Phi$. The physics basis for this relation is poorly understood.}
  %
  %
  {We use three-dimensional (3D) magnetohydrodynamics (MHD) numerical models of the coronae above active regions, that is, strong concentrations of magnetic field, to investigate the $L_{\rm{X}}$ versus $\Phi$ relation and illustrate this relation with an analytical model based on simple well-established scaling relations.}
  %
  %
  {In the 3D MHD model horizontal (convective) motions near the surface induce currents in the coronal magnetic field that are dissipated and heat the plasma. This self-consistently creates a corona with a temperature of 1 MK. We run a series of models that differ in terms of the (unsigned) magnetic flux at the surface by changing the (peak) magnetic field strength while keeping all other parameters fixed.}
  %
  %
  {In the 3D MHD models we find that the energy input into the corona, characterized by either the Poynting flux or the total volumetric heating,  scales roughly quadratically with the unsigned surface flux $\Phi$. This is expected from heating through field-line braiding. Our central result is the nonlinear scaling of the X-ray emission as $L_{\rm{X}}\propto \Phi^{3.44}$. This scaling is slightly steeper than found in recent observations that give power-law indices of up to only 2 or 3. Assuming that on a real star, not only the peak magnetic field strength in the active regions changes but also their number (or surface filling factor), our results are consistent with observations.}
  %
  %
  {Our model provides indications of what causes the steep increase in X-ray luminosity by four orders of magnitude from solar-type activity to fast
rotating active stars.}

  
\maketitle
%
%
\section{Introduction}
Stellar coronal X-ray emission is observed to increase with stellar rotation rate \citep[e.g.,][]{Pizzolato2003,Wright2011,Reiners2014, Magaudda2020}.
It is widely assumed that an increase in rotation could be responsible for stronger dynamo action leading to larger surface magnetic field.
Some active stars (e.g., M dwarfs) that rotate rapidly (typical periods of 1 to 2 days)  show high average photospheric magnetic field strengths which can reach up to 8 kG or even more \citep{Reiners}.
Because of this increased photospheric magnetic field, we can expect that a stronger upward-directed Poynting flux is generated that can heat the corona to higher temperatures, leading to stronger  X-ray emission in the corona.
An indication of such behavior has been found in stellar observations \citep[e.g.,][]{Vidotto} revealing a close relation between coronal X-ray emission and surface magnetic flux.

The scaling relationship between the coronal X-ray emission $L_{\rm{X}}$ and the surface magnetic flux $\Phi$ has been extensively studied using solar and stellar observations, and follows a power law, $L_{\rm{X}}\propto \Phi^{m}$.
In early studies, the power-law index $m$ was found to be close to unity \citep{Fisher,Pevtsov}, that is, the X-ray radiation scales almost linearly with magnetic flux.
However, more recent studies suggest a much steeper power law with $m=1.80$ \citep{Vidotto} or even steeper with $m=2.68$ \citep{Kochukhov}.
The physical mechanism relating the observed X-ray emission to the surface magnetic flux is still under debate.

To study the impact of the surface magnetic field on the coronal X-ray emission in the environment of a realistic setup, the use of 3D magnetohydrodynamic (MHD) models is required.
In addition, the 3D numerical simulations will provide a useful tool with which to further test the validity of a simplified analytical model.
The main advantage of 3D models is the self-consistent treatment of the  corona.
The heating originates from the Ohmic dissipation of currents induced by  photospheric magneto-convective motions. This  drives the magnetic field similar to Parker's field line braiding (or nanoflare) model \citep{parker1972,parker1983}.
The Parker field-line braiding model has been extensively studied in numerical models. The energy was shown to cascade in current sheets from large scales to dissipative scales before converting to heating \citep{Rappazzo2008}. 
In addition, 3D MHD simulations of footpoint motions  have proven successful in forming a self-consistent corona \citep[see e.g.,][]{GN02, GN05a, GN05b,Bingert2011, Bp13,Hansteen2015, Dahlburg2016, Dahlburg2018, Matsumoto2021}.

The numerical models are able to provide the necessary energy flux in the corona to heat it to temperatures beyond 1\,MK  and this energy flux is consistent with observations \citep{Bingert2011, Hansteen2015}. 
Furthermore, extreme ultraviolet (EUV) synthetic spectra from these 3D simulations can explain some aspects of the actual observations \citep{peter2004, Dahlburg2016, Warnecke2019}. This confirms the validity and efficiency of Parker's field line braiding model to create a hot corona.
These models can also be used to study the effects of magnetic helicity injection in the photosphere of active stars on the resulting coronal X-ray emission \citep{Warnecke2020}. 
This showed that an increase of photospheric magnetic helicity without changing the vertical magnetic field increases the coronal X-ray emission following simple power-law relations. 
However, the effect of the surface magnetic activity on the coronal X-ray emission in 3D MHD models of solar and stellar coronae has not yet been studied.
 
In our study, we focus on the effect that the photospheric magnetic field strength has on the coronal X-ray emission. 
This is motivated by the observation that stars more active than the Sun host a stronger surface magnetic field. 
For that reason, we choose to increase the strength of the vertical surface magnetic field at the bottom boundary of our computational domain; that is, we treat the peak (or average) magnetic field strength as a free parameter.
All the other parameters remain the same in all numerical experiments.
By varying only one parameter (i.e., the surface magnetic field) we can study the exact relation between magnetic  flux and coronal emission.
Other parameters important for the coronal energy input, such as the photospheric velocity distribution, are poorly constrained by observations for other stars and are therefore not changed (or varied) in the present work.
Our main objective is to relate the synthetic X-ray emission from the numerical models with the surface magnetic flux therein and compare this to the observed relationships. Furthermore, we compare our numerical results to our earlier analytical model \citep{Zhuleku}, which is briefly summarized in Sect.\,\ref{analytic}.

\section{Analytical scaling relations}\label{analytic}

The dependence of the coronal X-ray emission on the surface magnetic field has been extensively studied for the Sun as well as for other stars \citep{Fisher,Pevtsov,Vidotto,Kochukhov}. In \cite{Zhuleku}, we developed an analytical model to describe the $L_{\rm{X}}\propto \Phi^{m}$ relation, where $L_{\rm{X}}$ is the coronal X-ray emission and $\Phi$ the total surface unsigned magnetic flux. 

Our model is based on the well-known Rosner, Tucker \& Vaiana \cite[RTV;][]{Rosner} scaling laws.
These authors derived scalings relating the volumetric heating rate and the loop length to the coronal temperature and pressure.
Alternatively, we can express the RTV scaling laws in a way that allows us to relate temperature $T$ and number density $n$ with the volumetric heating rate $H$ and loop length $L$,
\begin{eqnarray}
\label{E:RTV.T}
T &\propto& H^{2/7}L^{4/7},
\\
\label{E:RTV.n}
n &\propto& H^{4/7}L^{1/7}.
\end{eqnarray}
Using these scaling laws together with other relations, we derive an analytical expression for the X-ray emission $L_{\rm{X}}$,
\begin{equation}\label{E:Lx.phi}
\begin{array}{@{}r@{~~}c@{~~}l}
L_{\rm{X}} &\propto& \displaystyle
               \Phi^m    \qquad \mbox{with}
\\[1ex]
m &=& \displaystyle
\frac{\beta\,\gamma}{7}\,\Big(2\alpha+8\Big)
  ~+~ \delta ~ \left(~ \frac{4}{7} 
                 + \frac{1}{7}\,\alpha
                 - \frac{8}{7}\,\beta\,\gamma
                 - \frac{2}{7}\,\alpha\,\beta\,\gamma
              ~\right).
\end{array}
\end{equation}
The power-law index $m$ depends only on four parameters $\alpha$, $\beta$, $\gamma$, and $\delta$.
The first, $\alpha,$ is related to the temperature sensitivity of the  instrument used for the X-ray observations.
In general, the X-ray radiation is proportional to the density squared and to a temperature-dependent function $R(T)$.
This function $R(T)$ is the sum of all the contribution functions of emission lines and continua and is also known as the temperature response function, and  differs from one instrument to the next.
We found that the temperature response function for temperatures below $\log_{10}T[K]\,{=}\,7$ can be expressed as a power law \citep[cf.][their Fig.1]{Zhuleku},
\begin{equation}\label{E.ana.R.vs.T}
    R\propto T^{\alpha}.
\end{equation}
The parameters $\beta$ and $\gamma$ characterize the relation between the energy flux, or the vertical Poynting flux $S_z$, injected in the corona and the vertical surface magnetic field $B$ and the volumetric heating rate $H$, again through power laws,
\begin{eqnarray}
S_{z}&\propto & B^{\,\beta} \label{E.ana.Sz.vs.B}
\\
H&\propto & S_{\!z}^{\,\gamma}. \label{E.ana.H.vs.Sz}
\end{eqnarray}
The case $\beta=1$ represents Alfv\'en wave heating and $\beta=2$ represents  nanoflare heating (see also Sect.\,\ref{S.heat}).
The parameter $\gamma$ was considered to be unity, $\gamma{=}1$.
Finally, $\delta$ relates the surface area covered by a magnetic structure (e.g. a whole active region) to the total magnetic flux,
\begin{equation}\label{A-Phi}
    A\propto \Phi^{\delta}
\end{equation}
Solar studies suggest a value of $\delta=0.819$ \cite[][]{Fisher}.

In our analytical study, we found the power-law index $m$ to be in the range from roughly one to almost two  \citep{Zhuleku}.
This result agrees relatively well with most observations, but at least one more recent observation finds an even steeper power-law connection of $L_{\rm{X}}{\propto}\Phi^{m}$ with $m$
of  2.68 \citep{Kochukhov}.

In the numerical study presented in the following sections, we assume that the area covered by the magnetic field remains the same, that is, the change of the surface magnetic flux is solely due to the (average) vertical magnetic field strength. This is equivalent to choosing $\delta{=}0$ in Eq.\,(\ref{A-Phi}).  In this case, we get a much steeper power law in the analytical model, with $m$ up to 4 \citep[][Sect.\, 5.2, Eq.\,15]{Zhuleku}. The numerical study in this paper provides a more detailed comparison to observations that closely match those of \cite{Kochukhov}.

\section{Numerical model setup}
 \subsection{Basic equations}
 
Our model is based on the work of \cite{Bingert2011,Bp13}.
We use the {\sc Pencil Code} \citep{PC2020}, where we numerically solve the MHD equations from the photosphere up to the corona.
The {\sc Pencil Code} is a finite difference code with a sixth-order numerical scheme used for spatial derivatives and third-order Runge-Kutta numerical scheme used for time derivatives.
The size of the computational box is 128$\times$128$\times$128 grid points in Cartesian coordinates ($x$, $y$, $z$), representing a 50$\times$50$\times$50 Mm$^{3}$ volume with a grid-scale of 390 km in all directions.
This comparably small grid size allows us to run a sufficiently large number of numerical experiments to study the relationship between the magnetic activity and coronal emission.

In principle, a higher resolution could affect the results, in particular because we are only barely resolving the granulation scale (on the Sun). However, earlier studies showed that reducing the grid spacing from around 550\,km by about a factor of four gives a similar overall appearance of the corona, although the finer details become visible \cite[][]{Chen2014,Chen2015}. 
Based on these findings we consider the results we get at the grid spacing of 390\,km to give a good representation of the injected photospheric Poynting flux then powering the hot corona.
This is also supported by earlier studies which used a grid spacing comparable to that of our model. At comparable resolution, \cite{GN05b, GN05a} found a loop-dominated active region. In a more recent data-driven model, \cite{Warnecke2019} used a grid spacing very close to ours (350\,km), albeit for a much larger computational domain, and found very good agreement between model and observations.

The MHD equations are the continuity, momentum, energy, and induction equation connecting density $\rho$, velocity $\boldsymbol{u}$, and temperature $T$ with the magnetic field $\boldsymbol{B}$, and pressure $p$:
 \begin{equation}
     \frac{{\rm D} \ln\rho}{{\rm D}t} +\boldsymbol{\nabla} \cdot \boldsymbol{u}=0, 
     \label{eq:cont}
     \end{equation}
     \begin{equation}
     \frac{{\rm D} \boldsymbol{u}}{{\rm D} t} =\frac{1}{\rho}\left[-\boldsymbol{\nabla} p+\rho \boldsymbol{g} +\boldsymbol{j} \times \boldsymbol{B}+2\nu\boldsymbol{\nabla} \circ \left(\rho \underline{S}\right)\right], \label{eq:mom}
     \end{equation}
     \begin{equation}
     \begin{array}{@{}l@{~~}}
         \displaystyle
         \frac{{\rm D} \ln T}{{\rm D}t}+(\gamma -1)\boldsymbol\nabla\cdot
\boldsymbol{u} ~~= 
\\
\displaystyle \hspace*{4em}
=~~\frac{1}{c_{V}\rho T}\left[\eta\mu_{0}\boldsymbol{j}^2+2\rho\nu\underline{S}^2+\boldsymbol{\nabla}
\cdot \boldsymbol{q}+L_{{\rm rad}}\right]. 
     \end{array}
    \label{eq:ener} 
 \end{equation}
Here the Lagrangian derivative ${\rm D}/{\rm D}t$ is defined as ${\rm D}/{\rm D}t=\partial/\partial t +\boldsymbol{u}\cdot \boldsymbol{\nabla}$. The current density is given by $\boldsymbol{j}=\boldsymbol{\nabla} \times \boldsymbol{B}/\mu_0$. 
The adiabatic index of an ideal gas is given by $\gamma=5/3$.
We use a constant gravitational acceleration, $\boldsymbol{g}=(0,0,-g),$ where $g=274$ m/s$^{2}$, $c_{V}$ is the specific heat capacity at constant volume, and $\underline{S}=(u_{i,j}+u_{j,i})/2-\delta_{ij}\boldsymbol{\nabla} \cdot \boldsymbol{u}/3$ is the rate of the strain tensor. 
The resistivity $\eta$ and viscosity $\nu$ are constants through the whole computational box with values $\eta=\nu=5\times 10^{10}$ m$^{2}$s$^{-1}$. At the bottom boundary we reduce $\eta$ to prevent excessive diffusion of the photospheric magnetic field. 

The magnetic field is calculated through the vector potential as,
 \begin{equation}
    \boldsymbol{B}=\boldsymbol{\nabla} \times (\boldsymbol{A}+ \boldsymbol{\nabla} \phi),
 \end{equation}
which ensures that the
$\boldsymbol{\nabla} \cdot \boldsymbol{B}=0$ is satisfied at all times. For our simulations we choose the resistive gauge $\phi=\eta \boldsymbol{\nabla}\cdot \boldsymbol{A}$ so that the induction equation reads,
 \begin{eqnarray}
    \frac{\partial \boldsymbol{A}}{\partial t}=\boldsymbol{u} \times \left(\boldsymbol{\nabla} \times \boldsymbol{A}\right)+\boldsymbol{\nabla}\left(\eta \boldsymbol{\nabla} \cdot \boldsymbol{A}\right).
 \end{eqnarray}

To close the system of equations we need the equation of state for an ideal gas which connects the gas pressure with temperature,
 \begin{eqnarray}
    p=\frac{k_{B}}{\mu\, m_{p}}\rho T,
 \end{eqnarray}
where $\rho$ is the mass density, $k_{B}$ is the Boltzmann constant, $\mu$ is the mean atomic weight and $m_{p}$ is the proton mass.

The radiative losses $L_{\rm{rad}}$, are modeled with an optically thin approximation using the radiative loss function $P(T)$. 
The function $P(T)$ depends on the abundances as well as the radiation processes that take place, such as the ionization rate and recombination rate \citep{Meyer,Murphy,Cook}.
For a detailed discussion on the implementation of the radiative loss function in the code, see \cite{Bingthesis}. 

The (Spitzer) heat flux along the magnetic field lines  $\boldsymbol{q}$ reads,

\begin{eqnarray}
\boldsymbol{q}=K_{0}T^{5/2}\,\hat{\boldsymbol{b}}(\hat{\boldsymbol{b}}\cdot \boldsymbol{\nabla} T),
\label{eq:heatflux}
\end{eqnarray}
where $K_{0}=10^{-11}$ W(mK)$^{-1}$  \citep{Spitzer} and $\hat{\boldsymbol{b}}$ is the unit vector of the magnetic field. 
To speed up the simulations, we replace Eq. \ref{eq:heatflux} by a nonFourier heat-flux scheme. See \cite{Warnecke} for a detailed description and discussion. 
Additionally, we use a semi-relativistic correction to the Lorentz force similar to the work of \cite{Boris1970APM}, \cite{Gombosi} and \cite{Rempel17}. For details of the implementation, see \cite{CP18} and \cite{Warnecke}.
 
One concern for all existing 3D models including heat conduction is the limited resolution in the transition region. Because the heat conductivity gets very inefficient for low temperatures (${\propto}T^{5/2}$), the temperature gradient has to become very steep in the transition region. This can necessitate a grid spacing along the magnetic field down to a km or less, which can be achieved in 1D models, in particular when using adaptive grids \cite[e.g.,][]{1993ApJ...402..741H} that then also allow tiny condensations to be resolved \cite[e.g.,][]{2003A&A...411..605M}.
When not having full resolution, this mainly leads to underestimation of the coronal density \cite[e.g.,][]{2013ApJ...770...12B}. This can be easily understood, because at low resolution the transition region gets smeared out and can radiate the downward conducted heat at a lower density, which also leads to lower density in the corona. In 3D models, such resolution is (currently) impossible, but there are ways to counteract these shortcomings \cite[e.g.,][]{2009ApJ...690..902L}. Despite their limited resolution, 3D MHD models can capture the essential response of the upper atmosphere to heat input and are consistent with the RTV scaling relations \cite[e.g.,][]{Bourdin}. In our models we are mostly concerned with the change in X-ray luminosity with magnetic activity, which depends on the change of density (or pressure) in the corona, but not on its absolute value. Indeed, we are likely to find that our numerical model underestimates the coronal density. Still, even if the absolute values of the density are not fully correct, we can still retrieve the proper scaling of density, and hence of the X-ray emission, which is the aim of our study.

In order to avoid instabilities because of steep gradients in temperature and density, we include in Eq. (\ref{eq:ener}) a (numerical) isotropic heat conduction term and a mass diffusion term \citep[see][]{Bingthesis}.
As is commonly used in many codes for modeling the solar corona \cite[see e.g.,][]{Gudiksen2011,Rempel17} the mass diffusion term is used to smooth out density fluctuations on the grid scale and has no effect on the coronal dynamics.
We also include a shock viscosity term for numerical reasons.
 
\subsection{Initial and boundary conditions}\label{S.init.boundary}
  
\begin{figure}
\centering
\includegraphics[width=\columnwidth]{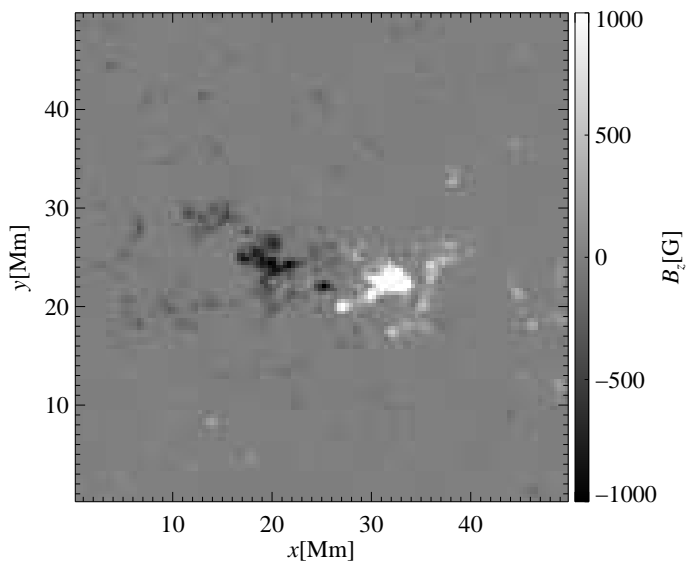}
\caption{Initial vertical magnetic field at the bottom boundary of the simulation. This is based on a magnetogram from  active region AR 11102 observed with HMI. See Sect.\,\ref{S.init.boundary}.}
\label{Figmag}
\end{figure}

The simulations are driven by (horizontal) motions on the solar surface that drive the magnetic field anchored in the photosphere. For the spatial distribution of the vertical magnetic field we employ an observed solar magnetogram (see \fig{Figmag}).
This is a snapshot of the active region AR 11102 as observed with the Helioseismic and Magnetic Imager \citep[HMI;][]{HMI} on August 30, 2010.
A detailed description of a similar active region and how it is implemented
in the model is discussed in \cite{Warnecke}. These kinds of magnetograms represent a typical situation for a solar active region and similar magnetograms have been used as input for data-driven simulations in earlier studies \citep[see e.g.,][]{GN05a, GN05b,Bingert2011}.

For the initial condition, we use a potential-field extrapolation to fill the box with magnetic field. 
The initial temperature stratification follows a  vertical profile mimicking the temperature increase into the solar corona. 
The initial density is calculated from hydrostatic equilibrium and the system is initially at rest with all velocity components set to zero.  
Both initial temperature and density stratifications are motivated by the model of \cite{Vernazza1981}.
   
In the horizontal $x$-$y$ plane, all variables are periodic. 
At the bottom boundary,  the temperature $T$ and density $\rho$ have fixed values whereas the horizontal velocities, $u_{x}$, $u_{y}$ have zero vertical gradients. 
The vertical velocity $u_{z}$ is set to satisfy the divergence-free condition $\boldsymbol{\nabla} \cdot \boldsymbol{u}=0$. 
We also prescribe a photospheric velocity driver which generates random photospheric motions; 
these are time-dependent and create a velocity distribution and spectrum similar to the one observed on the solar photosphere.
As a result, we shuffle the footpoints of the magnetic loops mimicking the solar photospheric flows in a similar way to, e.g., \cite{GN02, GN05a, GN05b}. 
At the top boundary, all velocity components are zero. 
To prevent any heat flux going in or out of the computational domain, the gradients of temperature and density are set to zero. 
The magnetic field is potential both for the bottom and the top boundary.

\section{Numerical experiments}
\label{S.exp}   
\subsection{Setup}

\begin{table}
\caption{Summary of numerical experiments.}
\label{table:runs} 
\centering 
\begin{tabular}{r r c }
\hline\hline
Run & a\tablefootmark{(1)} & $\Phi$ [Mx]\tablefootmark{(2)}\\  
\hline
   1B & 1 & $7.4\times 10^{20}$ \\ 
   2B & 2 & $1.5\times 10^{21}$  \\
   5B & 5 & $3.7\times 10^{21}$   \\
   10B & 10 & $7.4\times 10^{21}$  \\
   20B & 20 & $1.5\times 10^{22}$  \\ 
\hline                              
\end{tabular}
\tablefoot{
\tablefoottext{1}{The photospheric magnetic field strength is increased by the factor a.}
\tablefoottext{2}{ $\Phi$ is the total surface unsigned magnetic flux.}
}
\end{table}

The main idea of this work is to start with an active region hosting only a small amount of total (unsigned) magnetic flux and then run models with  increasing  magnetic flux.
Our aim is to study how the increase of the photospheric magnetic field strength contributes to the heating, and thus the X-ray emission of the solar or stellar corona. 
 
Here we report the results from a set of five numerical experiments.
The original total unsigned surface magnetic flux of AR11102 is around $7
\times 10^{20}$ Mx, which is a typical value of a small solar active region. %
We increase the flux by multiplying the $B_z$ component of the surface magnetic field by a constant ranging from 1 to 20 (see \tab{table:runs}).
The spatial structure of the magnetic field remains unchanged. In all cases, the total magnetic flux at the bottom boundary is zero, that is, the surface magnetogram is balanced.

The heating of the coronal loops originates from the dissipation of Poynting flux into heat.
More specifically, the conversion of the photospheric magnetic energy to thermal energy is due to the dissipation of the currents created by the random photospheric motions.
The stronger the magnetic field, the more Poynting flux reaches the corona, resulting in a higher temperature and X-ray emission. 

In our setup, the different total unsigned magnetic fluxes correspond to   peak values of the magnetic field ranging from 1 to 20 kG inside the spots; hence our naming convention in \tab{table:runs}.
The value of 20 kG is very high for a solar active region.
Recent observations of solar active regions measured maximum values of the magnetic field strength on the order of 8 kG \citep{Sebastian}.
However, this high value was observed in only very small area on the active region light bridge.
Typically the peak magnetic field strengths on solar active regions are on the order of 2\,kG to 3\,kG.
Therefore, of the numerical experiments listed in \tab{table:runs}, the 5B run can be considered to represent a typical solar active region, with a typical value for the total unsigned flux.
However, higher values of surface magnetic field could be a common feature of very active stars \citep{Reiners2014}, and so the runs 10B and 20B could be considered to be representative of more active stars.

The increase in the surface magnetic field is not the only way to increase the surface magnetic flux.
Alternatively, we could increase the horizontal extent of an active region
while keeping the peak magnetic field strength the same.
This will be the main focus of a future study.

\begin{figure*}
\sidecaption
\includegraphics[width=12cm]{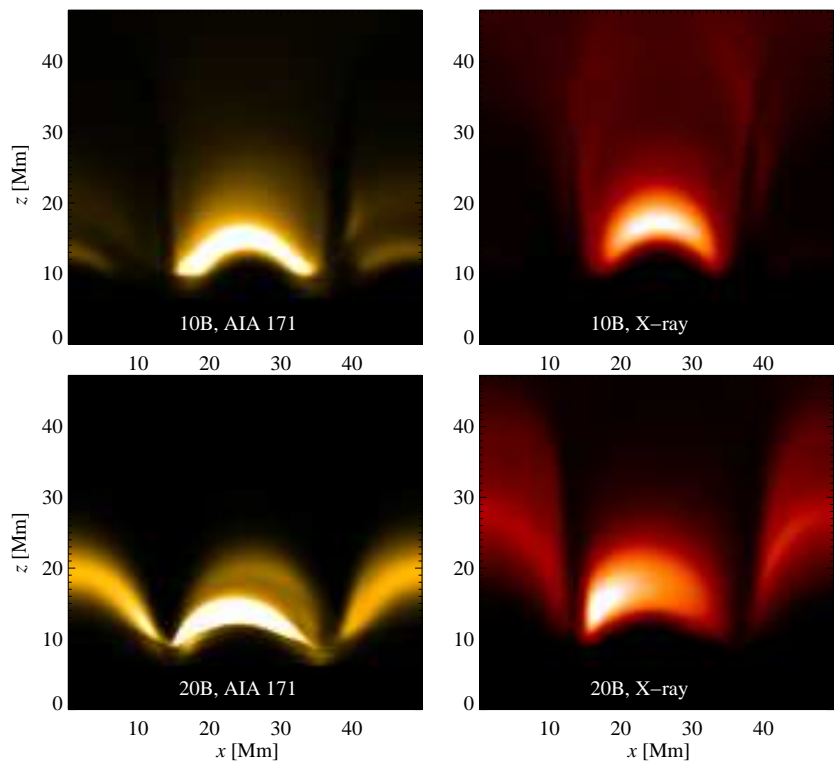}
\caption{Side view of the computational domain showing coronal emission. The left two panels show the emission as it would be seen by the 171\,{\AA} channel of AIA starting from around 1\,MK. The two right-hand panels show X-ray emission as seen by XRT sampling higher temperatures. Here we show snapshots of the two more active models, runs 10B and 20B. The emission here is integrated along the $y$ direction which corresponds to an observation near the limb (of the Sun or a star). The snapshot is taken at $t$=230 min; i.e., in the relaxed state. See Sect.\,\ref{S:heat.emiss}.}
\label{Figheat}%
\end{figure*}

\subsection{Synthesized emission: X-rays and EUV}\label{S:heat.emiss}

\begin{figure}
\centering
\includegraphics[width=\columnwidth]{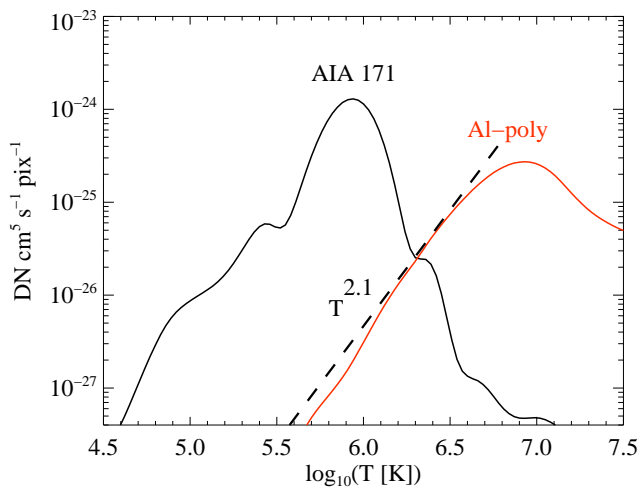}
\caption{Temperature response function for the AIA instrument on board SDO and the XRT on board Hinode. The black line shows the 171\,{\AA} channel of the AIA, the red line shows the Al-poly filter of the XRT. For illustration purposes, the black dashed line indicates a power-law approximation to XRT at temperatures below $10^7$\,K. See Sect.\,\ref{S:heat.emiss}.}
\label{response}
\end{figure}

Coronal loops are mostly observed in extreme ultraviolet (EUV) wavelengths and X-rays because these wavelengths give access to the 1 MK plasma in the corona, and in these wavelengths, the dilute corona is also visible in front of the solar (or stellar) disk, which is very bright in the visible range. Therefore, we synthesize the emission in one EUV and one X-ray band.
For the EUV band we choose the widely used 171\,{\AA} band as seen by  the Atmospheric Imaging Assembly \citep[AIA;][]{Lemen} on board
the Solar Dynamic Observatory \citep[SDO;][]{2012SoPh..275....3P}. 
Throughout the present paper we only consider the 171 {\AA} channel of the EUV spectra and not the whole EUV range.
We use the 171 {\AA} channel of AIA as a reference for plasma at 1 MK and how it compares with plasma at higher temperature emitting in X-rays.
For the X-ray emission, we use the Al-poly filter of the solar  X-ray
telescope \citep[XRT;][]{Golub2007} on board the Hinode observatory \citep{Kosugi}. 

As mentioned in Sect.\,\ref{analytic}, the optically thin radiative losses through lines and continua in
the corona observed in a given wavelength band  are given through
\begin{equation}\label{E:emissivity}
\varepsilon=n_{\rm{e}}^2\,R(T),
\end{equation}
where $n_{\rm{e}}$ is the electron density and $R(T)$ the temperature response (or contribution) function. The response needs to be calculated for each instrument (or filter) using the effective area depending on wavelength and the spectral lines forming in the wavelength region covered by the instruments. To calculate $R(T) $ for the 171\,{\AA} channel of AIA\ and the Al-poly filter of XRT, we use the routines  in the {\sc Chianti} data base v9 \cite[][]{1997A&AS..125..149D,2019ApJS..241...22D} as they are available in the {\sc SolarSoft} package\footnote{\url{http://www.lmsal.com/solarsoft}}. The response functions $R(T)$ for AIA 171\,{\AA} and XRT Al-poly are shown in \fig{response}. To calculate the synthetic images as they would be observed by AIA or XRT, $\varepsilon$ from Eq.\,(\ref{E:emissivity}) has to be integrated through the computational domain along the chosen line of sight. The samples for both channels for the 10B and 20B runs are shown in \fig{Figheat}. These are integrated along the $y$ direction which would correspond to an observation near the limb.

\subsection{Horizontal averages}\label{S:hor.avg}

The increase of the surface magnetic field results in an increase in temperature and density in the coronal part of the domain.
We calculate the horizontal averages of temperature $T$, density $\rho$, and the vertical component of the Poynting flux $S_{z}$  and then average these in time for an interval of 1 hour. During this time interval the computation reached a relaxed state, that is, the respective (spatially averaged) quantities show only rather small changes around a mean value (see \sect{S.S.temp.ev} and \fig{Figenergytime}). The average vertical stratification of $T$, $\rho$, and $S_z$ is shown in Fig. \ref{Figtemp}.

\subsubsection{Average Poynting flux deposited in the corona}

The photospheric horizontal motions lead to an upward-directed flux of magnetic energy, the Poynting flux.
Here we concentrate on its vertical component,
\begin{equation}
S_{z}=\eta(\boldsymbol{j}\times \boldsymbol{B})\big |_{z}-\frac{1}{\mu_{0}}(\boldsymbol{u}\times\boldsymbol{B}\times
\boldsymbol{B})\big |_{z}, 
\label{E.poyn}
\end{equation}which is shown in Fig. \ref{Figtemp}c for the different runs.
In the main part of the computational domain, the $\boldsymbol{u}{\times}\boldsymbol{B}{\times}\boldsymbol{B}$
term dominates and (on average) is positive, that is, upwards directed. The first term including the current, $\boldsymbol{j}$, is significant only near the bottom where boundary effects of the driving cause high currents. Energetically, this is not relevant, because there the density is high enough that the heating through the currents has virtually no effect.

As we increase the total unsigned magnetic flux from one experiment to the next, the energy
stored in  the corona increases. 
 The magnetic energy in excess of that of a potential field will be (partly) dissipated and converted into heat.
The higher amount of dissipated (free) magnetic energy in the runs with higher magnetic flux leads to  higher coronal temperatures and density (see \fig{Figtemp}a
and \fig{Figtemp}b). This is just as expected from the RTV scaling laws (Eqs.\,\ref{E:RTV.T} and \ref{E:RTV.n}) as we discuss later in Sect.\,\ref{rtv_num} (see also Fig.\,\ref{Figscalinggeneral}).

The Poynting flux for the runs 1B and 2B (i.e., black and blue solid lines in \fig{Figtemp}c) coincide in the lower part of the atmosphere (below 6 Mm).
We find this to be the case only for the specific time frame used for the time averaging (i.e., from 3.5 to 4.5 hrs).
For the time period before 3.5 hrs the Poynting fluxes for the runs 1B and 2B differ by a factor of two, as expected.
This peculiar behavior is only observed for the run 1B for which the magnetic field is too low to even produce a proper corona.
We observe this particular behavior only at the lower atmosphere while for the coronal part shown in \fig{Figenergytime}, we see a clear distinction between runs 1B and 2B.
However, as we only consider values on the coronal part for our results, this peculiar behavior will not have any effect on the results presented in the following section.

Furthermore, the Poynting flux of the 1B run at the base of the corona (i.e. at an average temperature of around 0.1 MK) barely reaches
50 W/m$^{2}$. 
Typical estimations based on observations suggest an energy requirement of around 100 W/$\rm{m}^{2}$
for the quiet Sun and  10$^4$ W/$\rm{m}^2$ above active regions \citep{withbroe}.
Therefore, we cannot expect this run with the lowest magnetic activity to produce a megakelvin corona.

 \subsubsection{Average temperature and density}
 
 \begin{figure}
\centering
\includegraphics[width=\columnwidth]{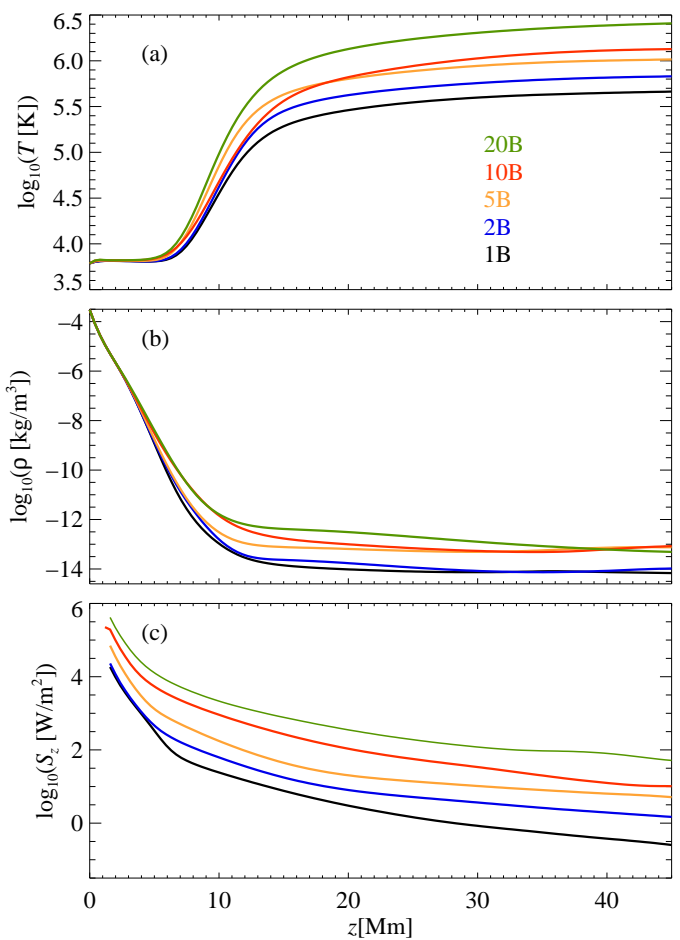}
\caption{Horizontal averaged quantities as a function of height. We show  temperature $T$ (panel a), density $\rho$ (panel b), and  the vertical component of the Poynting flux $S_{z}$ (panel c). The colors represent the different runs as indicated in the legend (cf. \tab{table:runs}). The quantities are averaged horizontally for each snapshot and then  in time  for 1\,hour  (between the time-point at 3.5 hr and that at 4.5 hr, as indicated in Fig.\,\ref{Figenergytime} by the vertical dashed lines. For the Poynting flux we omitted the first three grid points that show boundary effects. See Sect.\,\ref{S:hor.avg}.
}
\label{Figtemp}
\end{figure}
All of our simulations self-consistently form a hot upper atmosphere, where the temperature is about two orders of magnitude higher than at the surface (cf. Fig. \ref{Figtemp}a).
A higher total unsigned flux in the photosphere (cased 1B through 20B) corresponds not only to higher Poynting fluxes, but also to higher temperatures and density. The values shown in Fig. \ref{Figtemp}a and b are averages only, so the peak values are significantly higher, up to 5\,MK and more.

The experiment with the lowest magnetic activity (run 1B) fails to create a megakelvin hot corona, as expected. 
Still, we consider it in the analysis of the power-law relation in \sect{sec:scaling_num}.
The main focus of this work is to relate the coronal emission to the surface magnetic activity through a number of numerical experiments, and in this sense a model that is not active enough to produce a megakelvin corona also provides valuable insight.

Besides the increased temperature and density, the models with higher magnetic activity also have the transition region located at lower heights.
From this, it is clear that the height where the average temperature reaches 0.1 MK is lower %
for the runs with more magnetic flux. The higher energy input leads to a higher heat flux back to the Sun. Because the radiation is most efficient at lower temperatures (at 0.1 MK and below), in equilibrium the transition region will be found at lower temperatures and thus higher densities where it can radiate the energy. Consequently, the density (and the pressure) throughout the corona will be higher, as is indeed seen in our simulations.

The average density profile $\rho$ displays similar (qualitative) behavior to the temperature. In the coronal part, the density is high for the runs with higher magnetic flux (Fig.\,\ref{Figtemp}b). Following a steep drop over many orders of magnitude in the low atmosphere, the density remains almost constant in the coronal part. This is simply because of the large barometric (pressure) scale height at high temperatures. At 1\,MK, this scale height is about 50\,Mm and thus comparable to the vertical extent of our computational domain; hence the horizontally averaged pressure and density are roughly constant in the coronal part of our box.

\subsection{Temporal evolution}
\label{S.S.temp.ev}   
The quantities we consider in our model vary significantly with time, especially during the early phase of the simulations. To investigate the average behavior of our model, we have to consider a time-frame of the numerical model where the system reaches a relaxed (or evolved) state. During that state, the quantities show (comparably small) variations around an average value.

To illustrate this, we first consider the total heating in the coronal part of the computational domain. We define the coronal part as the volume above the height where the horizontally averaged
temperature is $10^5$\,K. 
Because the temperature gradient in the transition region around $10^5$\,K is rather steep, the exact choice of this temperature is inconsequential. We therefore define the total coronal heating $H_{\rm{tot}}$ as the volume integral over this coronal part, here symbolized by the subscript 'cor', 
\begin{equation}
{H_{\rm{tot}}}=\int_{\rm{cor}} \eta \mu_{0} \boldsymbol{j}^{2}dV.
\label{htot}
\end{equation}

\begin{figure}
\centering
\includegraphics[width=\columnwidth]{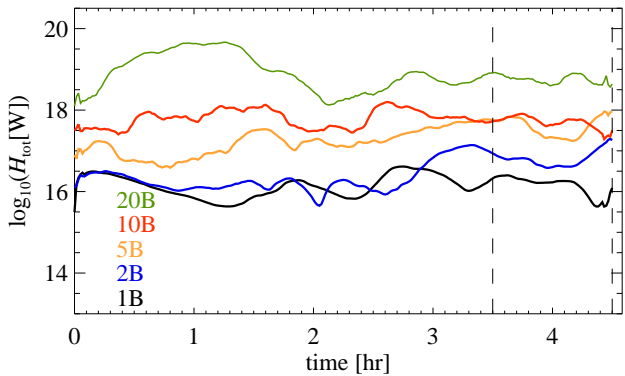}
\caption{Total coronal heating $H_{\rm{tot}}$ as a function of time. The vertical dashed lines indicate the time span used for the temporal averages.
The colors represent the different
runs as indicated in the legend (cf. \tab{table:runs}). See Sect.\,\ref{S.S.temp.ev}.}
\label{Figenergytime}
\end{figure}

The temporal variation of the heating is shown in 
Fig.\,\ref{Figenergytime} for the five models with different magnetic activity.
We see a clear ordering of the heating with the surface magnetic flux (increasing for run 1B through 20B), which we discuss in more detail in Fig.\,\ref{Figscaling} and Sect.\,\ref{S.heat}. In terms of the temporal evolution, we find that the heat input reaches a relaxed state rather quickly, probably within less than an hour.
This is expected, because the stresses applied on the magnetic field  in the photosphere will propagate with the Alfv\'en speed.
The corresponding Alfv\'en crossing time for perturbations to cross the whole box is on the order of minutes.

The situation for the relaxation time is different when considering the coronal emission in X-rays and the 171 {\AA} channel of AIA. Here it turns out that we have to wait for about three hours before the models reach a relaxed state. Mainly, this is because of the radiative cooling time under typical coronal
conditions which is on the order of 1 hour \citep{Aschwanden2006}.
Therefore, we examine the temporal evolution of the coronal radiation in some more detail below.

\begin{figure}
\centering
\includegraphics[width=\columnwidth]{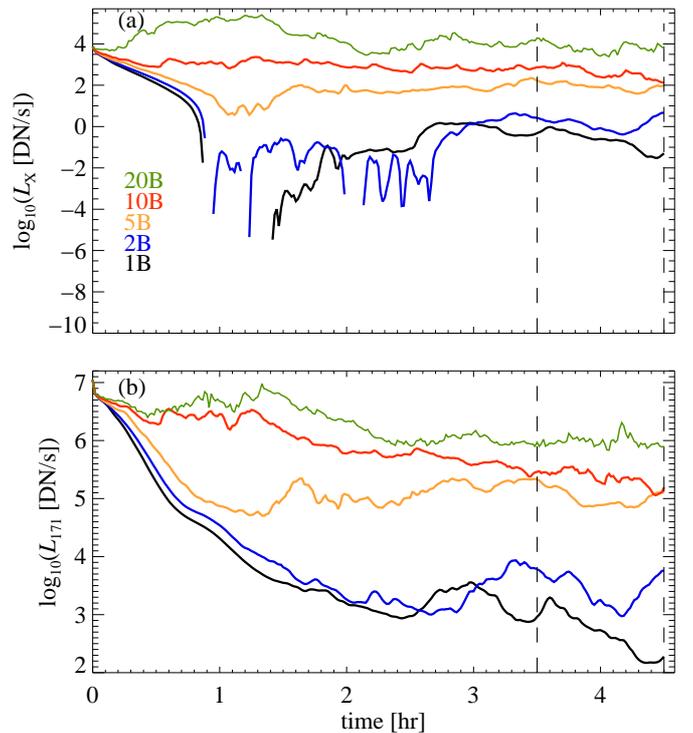}
\caption{Temporal evolution of the coronal emission from integration over the whole computational domain. Panel (a) shows the
X-ray emission  as seen by XRT in the Al-poly filter, panel (b) the EUV emission as it would be seen by AIA in the 171\,{\AA} channel. The vertical dashed lines
indicate the time-span used for time averaging. The colors represent the different
runs as indicated in the legend (cf. \tab{table:runs}). See Sect.\,\ref{S.S.temp.ev}.}
\label{Figemisstime}
\end{figure}

We now turn to the variability of the X-ray and EUV emission represented by the 171 {\AA} channel of AIA (see Fig. \ref{Figemisstime}).
Because the lower cool part of the atmosphere does not produce any significant amount of X-rays or EUV, we simply integrate the coronal emission over the whole computational domain. This is equivalent to the luminosity originating from the domain, $L_{\rm{X}}$ and  $L_{\rm{171}}$. Because we use the temperature response functions for XRT and AIA (Sect.\,\ref{S:heat.emiss}, Fig.\,\ref{response}), we get the counts per second as expected for the respective instrument from the whole loop in the box (cf. Fig.\,\ref{Figheat}). For the comparison between the different model runs, it is important that we use the same scale for the different models. Based on this, we see a clear scaling of the coronal emission with magnetic activity, in a similar way to what we see for the heat input. We discuss this in more detail in Fig.\,\ref{Figscalineuv} and Sect.\,\ref{Lx.vs.Phi}.

Overall, the coronal emission shows an initial drop on the timescale of almost an hour, in particular for EUV channel of AIA 171\,{\AA} in the runs with low magnetic activity. This is because the atmosphere of the initial condition is rather hot, and the plasma is cooling down until it reaches a new equilibrium after a few coronal radiative cooling times (Fig. \ref{Figemisstime}). Here we see that all model runs reach a relaxed state after about 2.5 to 3 hours. 

The relative variability for the 1B and 2B runs is larger than for the runs with higher magnetic activity. Still, the absolute variability seen in the more active runs is much higher (the logarithmic plot in \fig{Figemisstime}b shows the relative variation).
Also, when following run 1B further in time, the emission in the 171 {\AA} channel is not dropping further.
The situation is different when considering other channels like the X-ray regime which considers hotter plasma.
There, the variability is relatively low even for the low magnetic activity runs.
Thus, with some safety margin we can assume that from 3.5 to 4.5 hours the models have reached a relaxed state (see vertical dashed lines in Figs.\,\ref{Figenergytime} and \ref{Figemisstime}).
All the time averages discussed in our study are taken over this time frame.

\section{Scaling relations in numerical experiments}
\label{sec:scaling_num}

To characterize the model runs with different magnetic activity, that is, unsigned magnetic surface flux, we investigate the scaling relations in the form of power laws between different parameters.
In \sect{S.exp} we show that the enhancement of the photospheric magnetic flux leads to a substantial increase in temperature, density, Poynting flux, and coronal emission. 
In this section, we discuss the power-law relations of various quantities averaged in space and time quantities.
We discuss these results in Sect.\,\ref{S:dis} including the  study of \cite{Zhuleku}.

\begin{figure*}
\centering
\includegraphics[width=6 cm]{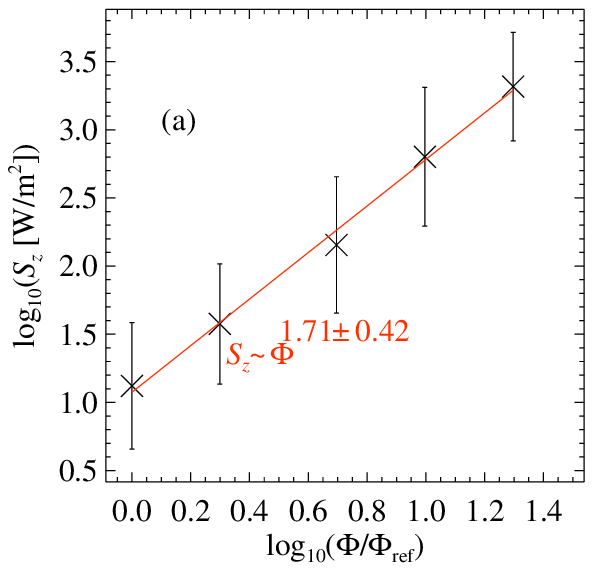}
\hspace{1.5cm}
\includegraphics[width=6 cm]{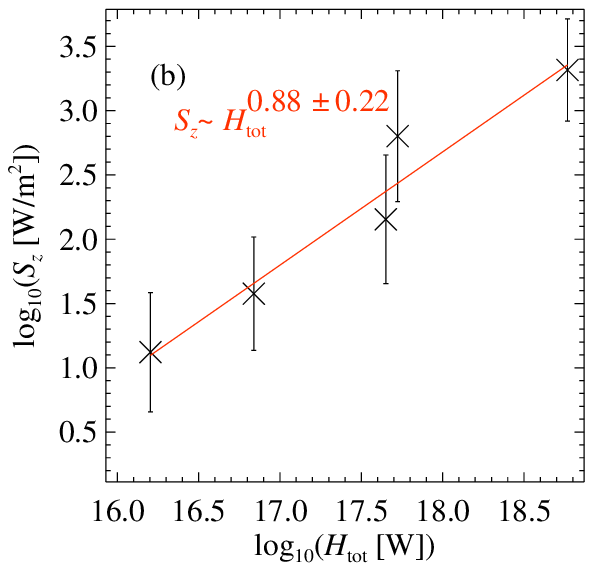}
\caption{Scaling  of average Poynting flux ${\langle}S_{z}{\rangle}$ with
unsigned surface flux $\Phi$ and average coronal heating ${\langle}H_{\rm
tot}{\rangle}$. Each data point represents an average for one of the model
runs with different unsigned surface magnetic flux as listed in \tab{table:runs}.
 As a reference for the magnetic flux, $\Phi_{\rm{ref}}$, we choose the magnetic
flux of the least active setup, run 1B (cf.\ \tab{table:runs}).  The bars
represent the standard deviation of $S_z$ in time. The red lines are power-law
fits to the data. See \sects{sec:scaling_num} and \ref{S.heat}.}
\label{Figscaling}
\end{figure*}

We first  concentrate on the relation between the vertical component of the average Poynting flux ${\langle}S_{z}{\rangle}$ and both the unsigned surface magnetic flux $\Phi$ and the averaged total coronal heating ${\langle}H_{\rm {tot}}{\rangle}$ (see \fig{Figscaling}).
Each cross in the two figures represents the average value of the respective quantities in each individual numerical model (cf. \tab{table:runs}). Here we take the horizontal average of the  Poynting flux ${\langle}S_{z}{\rangle}$ at the height where the horizontally averaged  temperature is $10^5$\,K, that is, the base of the corona and average it in time from 3.5 to 4.5\,hr (see \sect{S.S.temp.ev}).  
This represents the energy flux (per unit area) into the corona. The total averaged coronal volumetric heating ${\langle}H_{\rm {tot}}{\rangle}$ is calculated according to \eqn{htot} and then averaged between times 3.5 and 4.5\,hr as discussed in   \sect{S.S.temp.ev}. The unsigned surface flux $\Phi$ is the integral over the bottom boundary, that is, the stellar surface which is constant in time.
The bars in \fig{Figscaling} indicate the standard deviation of the respective quantity in time.
For relations displayed in \fig{Figscaling} we perform power-law fits (indicated by the red line) that result in
\begin{eqnarray}
  \label{E:num.S.vs.Phi}
  {\langle}S_{z}{\rangle} \propto \Phi^{\beta}~~~~~~~ &\quad \mbox{with} \quad & \beta~=1.71\pm0.42,
  \\
  \label{E:num.S.vs.H}
  {\langle}S_{z}{\rangle} \propto {\langle}H_{\rm{tot}}{\rangle}^{\Gamma} & \quad \mbox{with} \quad & \Gamma=0.88\pm0.22.
  \end{eqnarray}
Here $\Gamma$ corresponds to $1/\gamma$ from the analytical model in \eqn{E.ana.H.vs.Sz} and \cite{Zhuleku}.
The two scalings (\ref{E:num.S.vs.Phi}) and (\ref{E:num.S.vs.H}) imply that the  heating increases roughly quadratically with  magnetic flux, ${\langle}H_{\rm{tot}}{\rangle}\propto\Phi^{1.94}$.

As a next step, we relate the coronal temperature $T$ and density $\rho$ to the total coronal heating ${\langle}H_{\rm{tot}}{\rangle}$. 
Here we test to what extent the coronal temperature and density in our numerical model deviate from the analytic RTV scaling laws as given in \eqs{E:RTV.T} and (\ref{E:RTV.n}).
To this end, we calculate the average temperature ${\langle}T{\rangle}$ and density ${\langle}\rho{\rangle}$ in the corona in a height range from $z{=}10$\,Mm to 20\,Mm for each of the models.
 We choose this particular height range because this is where the bright structures appear (see \fig{Figheat}). If we were to also include higher regions of the box, the averages would no longer represent the visible parts of the corona. %
 In addition, we average ${\langle}T{\rangle}$ and ${\langle}\rho{\rangle}$ in time  from 3.5 to 4.5\,hr as discussed in \sect{S.S.temp.ev}.
We show the corresponding plots including the power-law fits of 
${\langle}T{\rangle}$ and ${\langle}\rho{\rangle}$ as a function of the total coronal heating $H_{\rm tot}$ in \fig{Figscalinggeneral}. 
The power-law fits yield
\begin{eqnarray}
  \label{E:num.T.vs.H}
   {\langle}T{\rangle} ~\propto~ {\langle}H_{\rm{tot}}{\rangle}^a & \quad \mbox{with} \quad & a=0.24\pm0.03,
\\
  \label{E:num.n.vs.H}
   {\langle}\rho{\rangle} ~\propto~ {\langle}H_{\rm{tot}}{\rangle}^b & \quad \mbox{with} \quad & b=0.76\pm0.05.
  \end{eqnarray}

Finally, we address the relation of the   X-ray emission to the heating rate and the unsigned surface magnetic flux. 
As the photospheric magnetic flux is larger for models with higher magnetic activity, the total  dissipated energy is also larger. Because the X-ray emission is expected to increase with the heating rate, it should also be larger for higher unsigned magnetic flux. For the analysis, we consider the averaged emission, ${\langle}L_{\rm{X}}{\rangle}$, that is integrated over the whole computational domain and averaged in the same way as the other quantities from 3.5 to 4.5\,hr. The corresponding relations are plotted in \fig{Figscalineuv} and \fig{Figscalineuvh}a, and the power-law fits give
\begin{eqnarray}
  \label{E:num.Lx.vs.Phi}
   {\langle}L_{\rm{X}}{\rangle} ~\propto~ \Phi^m\,~~~~~ & \quad \mbox{with}
\quad & m=3.44\pm0.28,
\\
  \label{E:num.Lx.vs.H}
   {\langle}L_{\rm{X}}{\rangle} ~\propto~ {\langle}H_{\rm{tot}}{\rangle}^q
& \quad \mbox{with}
\quad & \:q=1.78\pm0.15.
  \end{eqnarray}
We apply the same integration and time averaging to the EUV emission as seen by AIA in its 171\,{\AA} channel. Its relation to the heat input is displayed in  \fig{Figscalineuvh}b and the power-law fit reveals an almost linear relation,
\begin{equation}
  \label{E:num.EUV.vs.H}
   {\langle}L_{\rm{171}}{\rangle} ~\propto~ {\langle}H_{\rm{tot}}{\rangle}^p
 \qquad \mbox{with}
\qquad  p=1.05\pm0.09.
  \end{equation}

\section{Discussion}
\label{S:dis}

\subsection{Energy input into the corona}
\label{S.heat}

The solar coronal heating problem and the underlying physical mechanism has been extensively discussed over the last 70 years.
Two of the main mechanisms that are being considered are the Alfv\'en-wave model \citep[e.g.,][]{Balle2011} and the field-line braiding or nanoflare model \citep{parker1972,parker1983}.
In these two processes, the Poynting flux, and hence the heat input, have a different dependence on the magnetic field.  A simple estimate for these dependencies was given by  \cite{Fisher}, and here we follow their arguments. For an Alfv\'en wave with constant amplitude, the Poynting flux will be proportional to the propagation speed, that is,\ the  Alfv\'en velocity, and hence  to the magnetic field $B$, $S_{z}\propto B$. For the field-line braiding, the Poynting flux will be set by the driving motions with speed $u$ and the magnetic field, $S \propto u{\times}B{\times}B$, and hence the scaling will be $S_{z}\propto B^2$ (assuming that the driving motions do not change with $B$).

In our numerical experiments we find a power-law scaling between Poynting flux and (unsigned) magnetic flux with a power-law index of about $1.7{\pm}0.4$; see \eqn{E:num.S.vs.Phi} and \fig{Figscaling}a. Within the uncertainties this is consistent with the above (analytical) estimate of 2. This is not surprising, because our in our model we drive the coronal magnetic field through footpoint motions consistent with the field-line braiding scenario. Still, it is reassuring to recover this scaling by our numerical model.

Observationally, it is clear that Alfv\'en waves are present in the corona \citep[e.g.,][]{Tomczyk}. Still, it remains unclear as to whether or not the energy they carry is sufficient to energize the corona. Waves might play a role in the quiet Sun corona, but they appear to be unable to heat active regions \cite[][]{2011Natur.475..477M}.
Our particular model can contribute little to this discussion, because we do not fully resolve Alfv\'en waves, which is because of the comparably large dissipation.

Finally, we expect that the total energy dissipated in the coronal volume  matches the Poynting flux at the base of the corona. In this case, the Poynting flux should scale linearly with the total amount of energy dissipated in the corona. In our numerical experiments, we find a power-law relation with a power-law index of about $0.9{\pm}0.2$, which is consistent with linear; see \eqn{E:num.S.vs.H} and \fig{Figscaling}b.

\begin{figure*}
\centering
\includegraphics[width=6 cm]{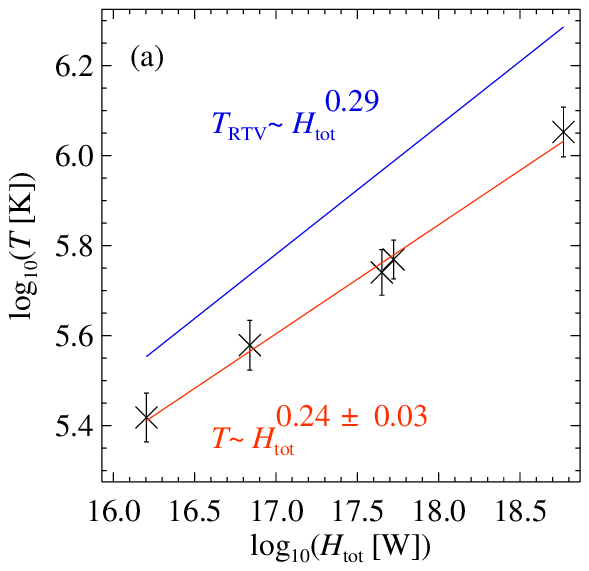}
\hspace{1.5cm}
\includegraphics[width=6. cm]{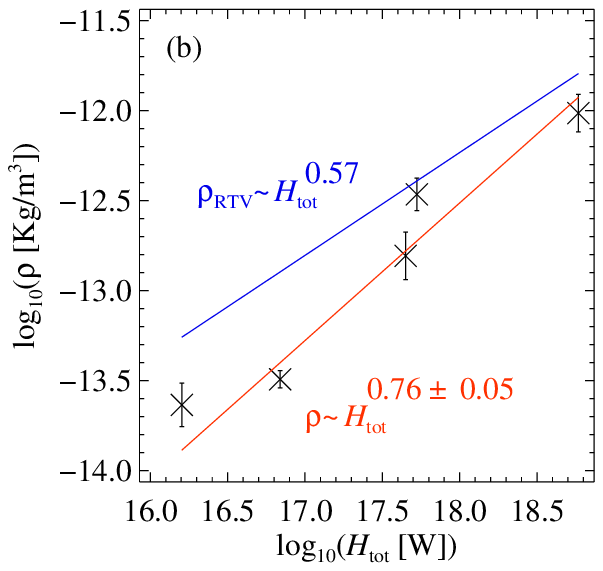}
\caption{Scaling of average temperature ${\langle}T{\rangle}$ and density
${\langle}\rho{\rangle}$ with  average coronal heat input ${\langle}H_{\rm
{tot}}{\rangle}$. Each data point represents an average for one of the model
runs with different unsigned surface magnetic flux listed in \tab{table:runs}.
The bars
represent the standard deviation of the spatial averages of $T$ and $\rho$
in time. The red lines are
power-law fits to the data.  The blue lines indicate what is expected from
the RTV scaling laws. See \sects{sec:scaling_num} and
\ref{rtv_num}.}
\label{Figscalinggeneral}
\end{figure*}

\subsection{RTV scaling laws compared to numerical experiments}\label{rtv_num}

One obvious check for the numerical experiments is to what extent the average quantities will follow the RTV scaling laws of \eqs{E:RTV.T} and (\ref{E:RTV.n}). Because of the spatial and temporal variability we cannot expect a perfect match, but the average quantities should roughly follow these scalings, as was found in an earlier model for one single (small) solar active region \citep{Bourdin}.

The original RTV scalings in \eqs{E:RTV.T} and (\ref{E:RTV.n}) include a dependence on the loop length $L$. 
However, in our numerical models the physical size of the computational box is kept constant.
Therefore, the coronal loops can be considered to have similar lengths.
Consequently, in this study we only have to consider the dependence of temperature and density on (total) heating in the coronal volume, $ H_{\rm{tot}}$.

For a comparison to the RTV scalings one should not only compare the power-law indices of the scaling, but also the absolute values of the temperature $T$ and density $\rho$ as given in the original paper by  \cite{Rosner}. To calculate the predictions from the RTV scalings, that is, $T_{\rm{RTV}}$ and $\rho_{\rm{RTV}}$, we use a loop length of  $L{=}30$ Mm, which is similar to the average
loop length we find in the emission patterns synthesized from the model (see \fig{Figheat}).
For the volumetric heating rate we use the  (total) heating in the coronal volume, ${\langle}H_{\rm{tot}}{\rangle}$, divided by the coronal volume, which gives the (average) volumetric heating.
The resulting variation of $T_{\rm{RTV}}$ and $\rho_{\rm{RTV}}$ with ${\langle}H_{\rm{tot}}{\rangle}$ is shown in \fig{Figscalinggeneral} (blue lines). The power law indices of 0.29 and 0.57 for these correspond to 2/7 and 4/7 in \eqs{E:RTV.T}
and (\ref{E:RTV.n}).

The power-law relation of the average temperatures ${\langle}T{\rangle}$
and density ${\langle}\rho{\rangle}$  in the numerical models with  ${\langle}H_{\rm{tot}}{\rangle}$ are close to what is expected from RTV. However, both the temperature and the density are significantly lower, by about a factor two and three; see \fig{Figscalinggeneral} and \eqs{E:num.T.vs.H} and (\ref{E:num.n.vs.H}).

The reason for this underestimation is mainly due to the averaging process of the temperature and density. The bright loops are hotter and denser than the ambient corona, meaning that the averages underestimate the values corresponding to the RTV scaling relations. Still, we conclude that the numerical models and the analytical scaling relations are consistent (as an order-of-magnitude estimation).

\subsection{Relation of X-ray emission to surface magnetic flux}\label{Lx.vs.Phi}

\begin{figure}
\centering
\includegraphics[width=6 cm]{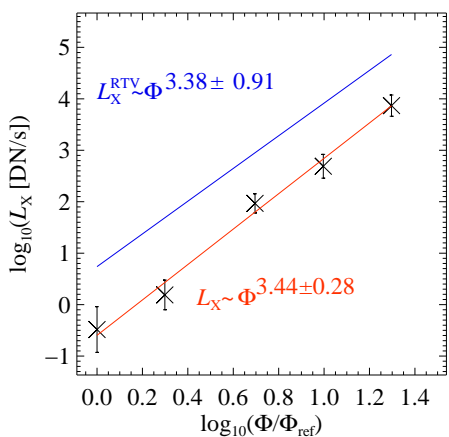}
\caption{Scaling of average X-ray emission ${\langle}L_{\rm{X}}{\rangle}$
with unsigned surface flux $\Phi$.  Each data point represents an average
for one of the model
runs with different unsigned surface magnetic flux listed in \tab{table:runs}.
 As a reference for the magnetic flux, $\Phi_{\rm{ref}}$, we choose the magnetic
flux of the least active setup, run 1B (cf.\ \tab{table:runs}).  The bars
represent the standard deviation of $L_{\rm{X}}$ in time. The red lines are
power-law fits to the data.
The blue lines show the analytic power-law relations based on the RTV scaling
laws;
see \sects{sec:scaling_num}, \ref{Lx.vs.Phi} and \ref{Lx.vs.Phi.ana}.
}
\label{Figscalineuv}
\end{figure}

The central result of our study is the power-law relation between the synthetic X-ray emission $L_{\rm {X}}$ and the unsigned surface magnetic flux $\Phi$.
We show this relation $L_{\rm{X}} \propto \Phi^{m}$ with a power-law index $m$ of about $3.4{\pm}0.3$ in \fig{Figscalineuv}; see also \eqn{E:num.Lx.vs.Phi}.

This $L_{\rm{X}}$ versus \ $\Phi^m$ relationship has been extensively discussed in numerous observational studies of the Sun and of stars of different spectral types \citep[see e.g.,][]{Fisher, Pevtsov, Vidotto}.
However, the physical reasons behind the observed power-law relation remain unclear.
Observations of various solar features and solar-like stars suggest a relation close to  but slightly steeper than linear, i.e., $m{\gtrsim}1$ \citep[e.g.,][]{Fisher,Pevtsov}.

We can estimate this power-law index $m$ from observations by comparing the (average) X-ray emission and magnetic field of the Sun to those of other stars.
The Sun has an average magnetic field of roughly 10 G \citep{Cranmer2017}, active stars show average fields of about 1 kG \citep{Saar1996, Saar2001}. This reveals a difference of a factor of 100 in magnetic field strength. 
At the same time the X-ray emission increases from the Sun to active stars by a factor of 1000 (X-ray-rotation activity relation) \citep[see e.g.,][]{Pizzolato2003}.
From this one can conclude that the $L_{\rm{X}}$ versus $\Phi$ increases as a power law with an of index $m=1.5$.
However, more recent studies find  the power law to be close to quadratic \citep[$m{=}1.8$;][]{Vidotto} or even steeper \citep[$m{=}2.68$;][]{Kochukhov}.
While \cite{Kochukhov} considered stars similar to the Sun in terms of spectral type, all these were considerably more active than the Sun in terms of X-ray luminosity.
The large sample of \cite{Vidotto} also mostly contained stars significantly more active than the Sun.

An almost linear relation of $L_{\rm{X}}$ versus \ $\Phi$ could be simply understood by increasing the number of active regions on a (solar-like) star. If the filling factor of active regions is low and one simply doubles the number of active regions, both the total unsigned magnetic flux on the stellar surface and the X-ray emission would double; hence, the linear relation between $L_{\rm{X}}$ and $\Phi$. This, in principle, could explain the findings of \cite{Fisher} and \cite{Pevtsov}.

\begin{figure*}
\centering
\includegraphics[width=6 cm]{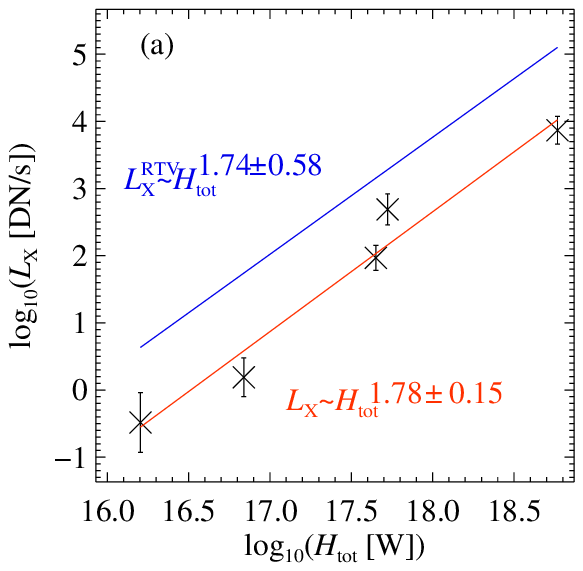}
\hspace{1.5cm}
\includegraphics[width=6 cm]{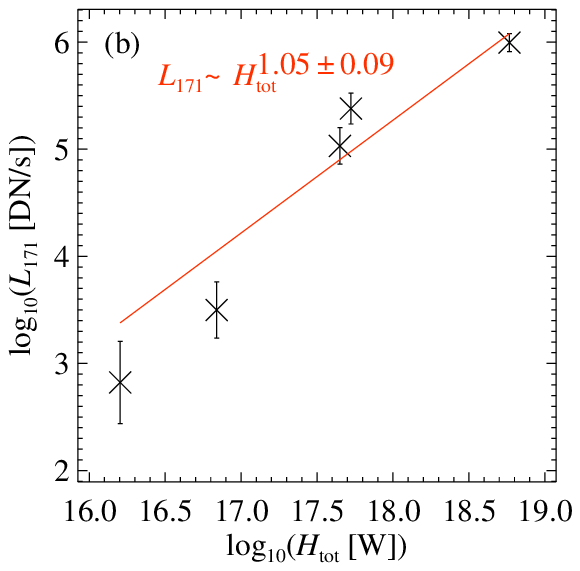}
\caption{Scaling of average X-ray and 171 {\AA} emission as observed by AIA, ${\langle}L_{\rm{X}}{\rangle}$,
and ${\langle}L_{\rm{171}}{\rangle}$
with
 average coronal heating ${\langle}H_{\rm
tot}{\rangle}$ for model
runs listed in \tab{table:runs}.  The bars
represent the standard deviation of $L_{\rm{X}}$
and $L_{\rm{171}}$ in time. The red lines are power-law fits to the data.
 The blue line in panel (a) indicates what is expected from
the RTV scaling laws. See \sects{sec:scaling_num} and \ref{S.X.EUV.vs.heat}.
}
\label{Figscalineuvh}
\end{figure*}

Active stars show X-ray luminosities (compared to their bolometric luminosity) that can be three or more  orders of magnitude larger than that of the Sun\citep[e.g.,][]{Vidotto}. In this case there would simply not be enough space on the star to cover it with enough (solar-like) active regions. This means that the X-ray emission per active region has to increase. This is exactly what we find in our model when increasing the magnetic flux of an active region while keeping its size (i.e. area) the same. This leads to a steep increase in X-ray luminosity with surface magnetic flux, an increase that is even steeper than observed: the power-law index $m$ for $L_{\rm{X}} \propto \Phi^{m}$ is about 3.4 in our model, while the largest value found in observations is  below 2.7 \cite[][]{Kochukhov}. This overestimation of the power-law index by our model indicates
that on real stars we might find a mixture of an increase in the numbers of active regions together with an increase in the peak magnetic field strength (or magnetic flux per active region) for more active stars.

Furthermore, the power-law index $m$ might be overestimated by our model.
Strong surface magnetic fields could quench the horizontal motions and consequently reduce the resulting Poynting flux \cite[e.g.,][]{GN05a}.
This is not included in our model.
In \app{A:quen}, we discuss the impact that this quenching could have on our estimations of the power-law relations.
We find that a realistic quenching of the horizontal velocities by about a factor of three would reduce the index $m$ of the power-law relation between X-ray emission and magnetic field, $L_{\rm{X}}\propto\Phi^m$ by about 25\% to around 2.6.
This would bring the value for $m$ derived from our model closer to the index derived by recent observations.

\subsection{Analytical model for scaling of X-ray emission}\label{Lx.vs.Phi.ana}

We now compare the scaling of $L_{\rm{X}} \propto \Phi^{m}$ to some basic analytical considerations. In an earlier study we derived an analytical scaling of the X-ray emission with surface magnetic flux that is based on the RTV scaling relations \cite[][]{Zhuleku}. As discussed in the summary of that model in \sect{analytic}, there we used a  scaling of active region size with (unsigned) magnetic flux based on solar observations, see Eq.\,(\ref{A-Phi}). In contrast, in our numerical model we keep the size or area covered by the active region constant; hence, here the power-law index in Eq.\,(\ref{A-Phi}) is $\delta{=}0$. This
simplifies our analytical scaling from \eqn{E:Lx.phi}
to
\begin{equation}
    m=\frac{\beta \gamma}{7}(2\alpha+8).
    \label{E.ms}
\end{equation}

According to \eqn{E.ana.R.vs.T}, $\alpha$ is the power-law index relating the temperature response (or contribution) function to temperature. Here we approximate this for X-rays as seen by the XRT on Hinode (see \fig{response}), where a power-law fit yields
\begin{equation}
   R_{\rm{X}}\propto T^\alpha \quad \mbox{with} \quad \alpha=2.1\pm0.2.
    \label{E.Rx.vs.T}
\end{equation}
We provide a more extensive discussion on the temperature responses for different instruments in \cite{Zhuleku}.

We take the other two parameters $\beta$ and $\gamma$ in \eqn{E.ms} from the relations of the Poynting flux to the unsigned surface magnetic flux  and the total heating in \eqs{E:num.S.vs.Phi} and (\ref{E:num.S.vs.H}), with $\gamma{=}1/\Gamma$ in \eqn{E:num.S.vs.H}.
This results in the analytic scaling (based on RTV) of 
\begin{equation}
    L_{\rm{X}}^{\rm{RTV}} \propto \Phi^{m'} \quad \mbox{with} \quad
    m'=3.38\pm0.91.
    \label{E.Lx.vs.Phi.ana.result}
\end{equation}
This is overplotted on the results from the numerical experiments in \fig{Figscalineuv} as a blue line.
We conclude that this  power-law index is in good agreement with the value we obtained from the numerical models, see \eqn{E:num.Lx.vs.Phi} and \fig{Figscalineuv}.

Just as for the comparison to the RTV scalings in \sect{rtv_num}, also here it is not sufficient to find a match of the power-law index; the absolute values of the derived X-ray emission also have to be of the same order for a good match. To get the constant of proportionality in \eqn{E.Lx.vs.Phi.ana.result}, we have to assign a volume of the emitting structure described in the analytical scaling. For the comparison to our numerical model we therefore assign the volume of the loop(s) dominating the coronal emission. Using the loops in \fig{Figheat} as a guideline, these have a length and a radius of about $L{\approx}30$\,Mm and $r{\approx}2.5$\,Mm. With the volume of
$V=\pi R^2L$ we then find the total X-ray radiation from the loops based on the (RTV) scaling relations of $L_{\rm{X}}^{\rm{RTV}}$. Within an order of magnitude these match what we find in the numerical study. The deviation is mainly because of the factor of about three difference in density (cf \sect{rtv_num}) which enters the emission quadratically.

\subsection{X-ray and EUV emission versus coronal energy input}\label{S.X.EUV.vs.heat}
Our current understanding of solar physics would lead us to believe that a higher energy input into the corona should result in a higher X-ray emission. Indeed, this is what we find in \fig{Figscalineuvh}a. However, we find an almost 
 quadratic relationship,
 ${\langle}L_{\rm{X}}{\rangle} \propto {\langle}H_{\rm tot}{\rangle}^{1.8}$, cf.\ \eqn{E:num.EUV.vs.H}.
This nonlinearity is counter-intuitive: 
Assuming that the plasma in our numerical model has reached a relaxed state, then whatever energy reaches the corona should be radiated away as X-ray emission (i.e., naively, we would expect a linear dependence).

This (roughly) quadratic relation can be understood when going back to the RTV scalings. The X-ray emission is given by $L_{\rm{X}}\propto n^2\,R_{\rm{X}}$. Combining this with the fit to the X-ray response function in \eqn{E.Rx.vs.T}
and the RTV\ scaling relations in \eqs{E:RTV.T}
and (\ref{E:RTV.n})
provides the analytical scaling for the X-ray emission with (average) coronal heat input,
\begin{equation}
    L_{\rm{X}}^{\rm{RTV}} \propto  {\langle}H_{\rm{tot}}{\rangle}^{q'} \quad \mbox{with} \quad
    q'=1.74\pm0.58.
    \label{E.Lx.vs.H.ana.result}
\end{equation}
This is shown in comparison to the results from the numerical model in \fig{Figscalineuvh}a as a blue line. As before, we assume a constant loop length, because the active region size is the same for all numerical experiments.

This consideration provides an understanding of the nonlinear relation between X-ray output and heat input. In the corona the energy input will be only partly balanced by the X-ray output (heat conduction, wind outflow, and radiation at other wavelengths also play their respective roles). Because the X-ray output shows a nonlinear temperature dependence, naturally it will also be connected in a nonlinear fashion to the heat input. The simple RTV scaling relations, together with the temperature response of X-ray emission give rise to the roughly quadratic dependence of X-rays on heat input.

The situation changes when investigating another wavelength region, which essentially implies probing a different temperature range. For this we use the channel as seen by AIA at 171\,{\AA}. The emission from the 171 {\AA} channel of AIA in this band mostly originates from temperatures around 1\,MK (cf. \fig{response}).
If we apply the same analysis as for the X-ray emission, the numerical experiments show an almost linear scaling with heat input, 
 ${\langle}L_{\rm{171}}{\rangle} \propto {\langle}H_{\rm tot}{\rangle}^{1.1}$; see \fig{Figscalineuvh}b. The reason this relation is less steep than for the X-rays is because of the different temperature response in the EUV.

The response of the 171 {\AA} channel of AIA cannot be simply approximated by a power law (see \fig{response}). However, over the temperature range of most our models, as a very rough zeroth-order approximation we could assume the response in the 171 {\AA} channel to be constant.
This would be consistent with the power-law index for the response function, that is, \ $\alpha$ in  \eqn{E.ana.R.vs.T}, being equal to zero. Following the above procedure for the X-rays, in analogy to \eqn{E.Lx.vs.H.ana.result} we would find $ L_{\rm{171}}^{\rm{RTV}} \propto {\langle}H_{\rm{tot}}{\rangle}^{1.14}$ for the 171 {\AA} channel emission.
This result compares well with the numerical experiments (cf.\ \fig{Figscalineuvh}b), but certainly it has to be taken cum grano salis, because of our very rough assumption of $\alpha{=}0$. This is why we do not overplot an analytical approximation in \fig{Figscalineuvh}b as we did for the other scaling relations. Still, this consideration underlines the importance of the temperature response in establishing a relationship between coronal emission and heat input.

\subsection{Relating X-rays to EUV emission}\label{S.X.vs.EUV}

\begin{figure}
\centering
\includegraphics[width=6 cm]{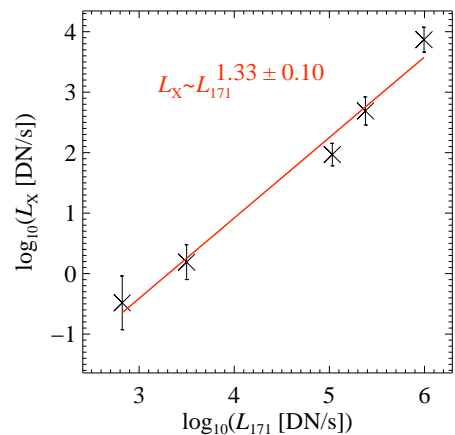}
\caption{ Scaling  of averages of X-ray emission ${\langle}L_{\rm{X}}{\rangle}$
with 171 {\AA} emission as observed by AIA ${\langle}L_{\rm
171}{\rangle}$ for model
runs listed in \tab{table:runs}.  The bars
represent the standard deviation of $L_{\rm{X}}$ in time. The red line
is a power-law fit to the data;
see \sect{S.X.vs.EUV}}
\label{Figscalineuvlx}
\end{figure}

Finally, we briefly investigate the scaling relation between the X-ray  and the emission of the 171 {\AA} channel of AIA. 
Solar and stellar studies have addressed the relation between the radiative fluxes in different wavelength bands, also called flux-flux relations.
 One prominent example is the relation between the coronal X-ray radiative flux and the radiance in the chromospheric \ion{Ca}{II} line with $F_{\rm{X}}\propto (F_{\rm{Ca\,II}})^{1.67}$ \citep[e.g.,][]{1983A&A...127..289S}. 
Power-law relations have also been derived with other chromospheric and transition region lines, such as
 \ion{Mg}{II}, \ion{Si}{II}, \ion{C}{II}, \ion{and C}{IV} \citep{Schrijver1987}.
The higher the temperature in the source region of the emission, the steeper the power-law relation of the respective line to the surface magnetic field \cite[e.g.,][]{2015RSPTA.37340259T}. This implies that the relation of ${\langle}L_{\rm{X}}{\rangle}$ versus ${\langle}L_{\rm 171}{\rangle}$ is less steep than the relation of X-rays to \ion{Ca}{II}.  

In the numerical models we find that the relation between X-rays and the 1\,MK EUV emission of the 171 {\AA} channel of AIA follows a power law with about ${\langle}L_{\rm{X}}{\rangle} \propto {\langle}L_{\rm 171}{\rangle}^{1.3}$; see \fig{Figscalineuvlx}. This is also consistent with other numerical studies \citep[see e.g.,][their Fig. 10b]{Warnecke2020}. Therefore, as expected, it is less steep than X-rays
to \ion{Ca}{II} with a power-law index of about 1.7.

Our numerical models are not realistic enough when it comes to the cooler parts of the atmosphere, in the transition region around $10^5$\,K and in particular in the cooler chromosphere. Still we see a consistent trend for the flux-flux relations, and their investigation is left for future more realistic models.

\section{Conclusions}

We performed a series of 3D MHD numerical simulations of active regions. With these, we address the question of how the (X-ray) emission from stellar coronae scales with the surface magnetic flux. Observationally this power-law scaling $L_{\rm{X}}\propto \Phi^{m}$ is well established with the most recent studies giving power-law indices $m$ in the range of below 2 to almost 3.
\citep{Vidotto,Kochukhov}.
So far, the physical basis for this relation is poorly understood in terms of (numerical) models. 

For our model series, we assumed that the area covered by the active region remains the same, and we changed the peak (or average) vertical field strength $B$ to change the (unsigned) magnetic flux at the surface by a factor of 20. This resulted in a change of the X-ray emission by more than four orders of magnitude. The scaling we found in our numerical experiments is a power law with an index $m\approx3.4$, which is slightly steeper than found in observations. As we discussed  in \sect{Lx.vs.Phi}, this difference can be understood if on the real star there are not only active regions with higher magnetic flux (but the same area) but also a larger filling factor of active regions, or in other words a larger number of active regions.

The results of our numerical experiments are consistent with an analytical scaling model \cite[][]{Zhuleku}. This is based on the RTV scaling laws connecting the temperature and density to the heating rate and size of the coronal structure and the temperature response of the wavelength band in which the observations are performed.
With this, we have a clear understanding of how and why the radiative X-ray output changes in a nonlinear fashion in response to the heat input, and hence the surface magnetic flux.

In our numerical model we choose the specific approach of increasing the surface magnetic flux by increasing the field strength. This is motivated by the very high field strengths seen on active M dwarfs of up to 8\,kG on average \cite[e.g.,][]{Reiners}. Such a large average suggests that the highest field strengths on these stars could be even  higher, perhaps consistent with our model. Still, to investigate further possibilities we will study the response of the coronal emission to an increase in the magnetic flux by increasing the area of the active region(s) in a future project.

Our models for individual active regions cannot be expected to fully account for all
aspects of the scaling of coronal emission with (unsigned) surface magnetic flux. Still, our approach indicates a way in which to  understand the excessive increase in the observed
X-ray emission by four orders of magnitude from solar-type activity to fast rotating active stars \citep[by e.g.,][]{Pizzolato2003,Vidotto}.

\begin{acknowledgements}
The simulations have been carried out at GWDG in G{\"o}ttingen and the Max Planck Computing and Data Facility (MPCDF) in Garching. This work was supported by the International Max Planck Research School (IMPRS) for Solar System Science at the University of G{\"o}ttingen. J.W.\ acknowledges funding by the Max-Planck-Princeton Center for Fusion and Astro Plasma Physics. This project has received funding from the European Research Council (ERC) under the European Union's Horizon 2020 research and innovation programme (grant agreement n:o 818665 ``UniSDyn'').
\end{acknowledgements}


\bibliographystyle{aa}
\bibliography{paper}

\begin{thebibliography}{65}
\expandafter\ifx\csname natexlab\endcsname\relax\def\natexlab#1{#1}\fi

\bibitem[{{Aschwanden}(2005)}]{Aschwanden2006}
{Aschwanden}, M.~J. 2005, {Physics of the Solar Corona. An Introduction with
  Problems and Solutions (2nd edition)} ({Springer-Verlag Berlin Heidelberg New
  York})

\bibitem[{Bingert(2009)}]{Bingthesis}
Bingert, S. 2009, Heating of the corona in a 3D MHD forward model approach
  (Univ. Freiburg (Breisgau)), \url {http://dx.doi.org/10.23689/fidgeo-19}

\bibitem[{{Bingert} \& {Peter}(2011)}]{Bingert2011}
{Bingert}, S. \& {Peter}, H. 2011, \aap, 530, A112

\bibitem[{{Bingert} \& {Peter}(2013)}]{Bp13}
{Bingert}, S. \& {Peter}, H. 2013, \aap, 550, A30

\bibitem[{Boris(1970)}]{Boris1970APM}
Boris, J.~P. 1970, in A Physically Motivated Solution of the Alfven Problem:NRL
  memorandum report

\bibitem[{{Bourdin} {et~al.}(2016){Bourdin}, {Bingert}, \& {Peter}}]{Bourdin}
{Bourdin}, P.-A., {Bingert}, S., \& {Peter}, H. 2016, A\&A, 589, A86

\bibitem[{{Bradshaw} \& {Cargill}(2013)}]{2013ApJ...770...12B}
{Bradshaw}, S.~J. \& {Cargill}, P.~J. 2013, \apj, 770, 12

\bibitem[{{Castellanos Dur{\'a}n} {et~al.}(2020){Castellanos Dur{\'a}n},
  {Lagg}, {Solanki}, \& {van Noort}}]{Sebastian}
{Castellanos Dur{\'a}n}, J.~S., {Lagg}, A., {Solanki}, S.~K., \& {van Noort},
  M. 2020, \apj, 895, 129

\bibitem[{{Chatterjee}(2020)}]{CP18}
{Chatterjee}, P. 2020, Geophysical and Astrophysical Fluid Dynamics, 114, 213

\bibitem[{{Chen} \& {Peter}(2015)}]{Chen2015}
{Chen}, F. \& {Peter}, H. 2015, \aap, 581, A137

\bibitem[{{Chen} {et~al.}(2014){Chen}, {Peter}, {Bingert}, \&
  {Cheung}}]{Chen2014}
{Chen}, F., {Peter}, H., {Bingert}, S., \& {Cheung}, M.~C.~M. 2014, \aap, 564,
  A12

\bibitem[{{Cook} {et~al.}(1989){Cook}, {Cheng}, {Jacobs}, \&
  {Antiochos}}]{Cook}
{Cook}, J.~W., {Cheng}, C.-C., {Jacobs}, V.~L., \& {Antiochos}, S.~K. 1989,
  \apj, 338, 1176

\bibitem[{{Cranmer}(2017)}]{Cranmer2017}
{Cranmer}, S.~R. 2017, \apj, 840, 114

\bibitem[{{Dahlburg} {et~al.}(2016){Dahlburg}, {Einaudi}, {Taylor},
  {Ugarte-Urra}, {Warren}, {Rappazzo}, \& {Velli}}]{Dahlburg2016}
{Dahlburg}, R.~B., {Einaudi}, G., {Taylor}, B.~D., {et~al.} 2016, \apj, 817, 47

\bibitem[{{Dahlburg} {et~al.}(2018){Dahlburg}, {Einaudi}, {Ugarte-Urra},
  {Rappazzo}, \& {Velli}}]{Dahlburg2018}
{Dahlburg}, R.~B., {Einaudi}, G., {Ugarte-Urra}, I., {Rappazzo}, A.~F., \&
  {Velli}, M. 2018, \apj, 868, 116

\bibitem[{{Dere} {et~al.}(2019){Dere}, {Del Zanna}, {Young}, {Landi}, \&
  {Sutherland}}]{2019ApJS..241...22D}
{Dere}, K.~P., {Del Zanna}, G., {Young}, P.~R., {Landi}, E., \& {Sutherland},
  R.~S. 2019, \apjs, 241, 22

\bibitem[{{Dere} {et~al.}(1997){Dere}, {Landi}, {Mason}, {Monsignori Fossi}, \&
  {Young}}]{1997A&AS..125..149D}
{Dere}, K.~P., {Landi}, E., {Mason}, H.~E., {Monsignori Fossi}, B.~C., \&
  {Young}, P.~R. 1997, \aaps, 125, 149

\bibitem[{{Fisher} {et~al.}(1998){Fisher}, {Longcope}, {Metcalf}, \&
  {Pevtsov}}]{Fisher}
{Fisher}, G.~H., {Longcope}, D.~W., {Metcalf}, T.~R., \& {Pevtsov}, A.~A. 1998,
  \apj, 508, 885

\bibitem[{{Golub} {et~al.}(2007){Golub}, {Deluca}, {Austin}, {Bookbinder},
  {Caldwell}, {Cheimets}, {Cirtain}, {Cosmo}, {Reid}, {Sette}, {Weber},
  {Sakao}, {Kano}, {Shibasaki}, {Hara}, {Tsuneta}, {Kumagai}, {Tamura},
  {Shimojo}, {McCracken}, {Carpenter}, {Haight}, {Siler}, {Wright}, {Tucker},
  {Rutledge}, {Barbera}, {Peres}, \& {Varisco}}]{Golub2007}
{Golub}, L., {Deluca}, E., {Austin}, G., {et~al.} 2007, \solphys, 243, 63

\bibitem[{Gombosi {et~al.}(2002)Gombosi, Tóth, Zeeuw, Hansen, Kabin, \&
  Powell}]{Gombosi}
Gombosi, T., Tóth, G., Zeeuw, D., {et~al.} 2002, Journal of Computational
  Physics, 177, 176

\bibitem[{{Gudiksen} {et~al.}(2011){Gudiksen}, {Carlsson}, {Hansteen}, {Hayek},
  {Leenaarts}, \& {Mart{\'\i}nez-Sykora}}]{Gudiksen2011}
{Gudiksen}, B.~V., {Carlsson}, M., {Hansteen}, V.~H., {et~al.} 2011, \aap, 531,
  A154

\bibitem[{{Gudiksen} \& {Nordlund}(2002)}]{GN02}
{Gudiksen}, B.~V. \& {Nordlund}, {\AA}. 2002, \apjl, 572, L113

\bibitem[{{Gudiksen} \& {Nordlund}(2005{\natexlab{a}})}]{GN05b}
{Gudiksen}, B.~V. \& {Nordlund}, {\AA}. 2005{\natexlab{a}}, \apj, 618, 1031

\bibitem[{{Gudiksen} \& {Nordlund}(2005{\natexlab{b}})}]{GN05a}
{Gudiksen}, B.~V. \& {Nordlund}, {\AA}. 2005{\natexlab{b}}, \apj, 618, 1020

\bibitem[{{Hansteen}(1993)}]{1993ApJ...402..741H}
{Hansteen}, V. 1993, \apj, 402, 741

\bibitem[{{Hansteen} {et~al.}(2015){Hansteen}, {Guerreiro}, {De Pontieu}, \&
  {Carlsson}}]{Hansteen2015}
{Hansteen}, V., {Guerreiro}, N., {De Pontieu}, B., \& {Carlsson}, M. 2015,
  \apj, 811, 106

\bibitem[{{Kochukhov} {et~al.}(2020){Kochukhov}, {Hackman}, {Lehtinen}, \&
  {Wehrhahn}}]{Kochukhov}
{Kochukhov}, O., {Hackman}, T., {Lehtinen}, J.~J., \& {Wehrhahn}, A. 2020,
  \aap, 635, A142

\bibitem[{{Kosugi} {et~al.}(2007){Kosugi}, {Matsuzaki}, {Sakao}, {Shimizu},
  {Sone}, {Tachikawa}, {Hashimoto}, {Minesugi}, {Ohnishi}, {Yamada}, {Tsuneta},
  {Hara}, {Ichimoto}, {Suematsu}, {Shimojo}, {Watanabe}, {Shimada}, {Davis},
  {Hill}, {Owens}, {Title}, {Culhane}, {Harra}, {Doschek}, \& {Golub}}]{Kosugi}
{Kosugi}, T., {Matsuzaki}, K., {Sakao}, T., {et~al.} 2007, \solphys, 243, 3

\bibitem[{{Lemen} {et~al.}(2012){Lemen}, {Title}, {Akin}, {Boerner}, {Chou},
  {Drake}, {Duncan}, {Edwards}, {Friedlaender}, {Heyman}, {Hurlburt}, {Katz},
  {Kushner}, {Levay}, {Lindgren}, {Mathur}, {McFeaters}, {Mitchell}, {Rehse},
  {Schrijver}, {Springer}, {Stern}, {Tarbell}, {Wuelser}, {Wolfson}, {Yanari},
  {Bookbinder}, {Cheimets}, {Caldwell}, {Deluca}, {Gates}, {Golub}, {Park},
  {Podgorski}, {Bush}, {Scherrer}, {Gummin}, {Smith}, {Auker}, {Jerram},
  {Pool}, {Soufli}, {Windt}, {Beardsley}, {Clapp}, {Lang}, \&
  {Waltham}}]{Lemen}
{Lemen}, J.~R., {Title}, A.~M., {Akin}, D.~J., {et~al.} 2012, \solphys, 275, 17

\bibitem[{{Lionello} {et~al.}(2009){Lionello}, {Linker}, \&
  {Miki{\'c}}}]{2009ApJ...690..902L}
{Lionello}, R., {Linker}, J.~A., \& {Miki{\'c}}, Z. 2009, \apj, 690, 902

\bibitem[{{Magaudda} {et~al.}(2020){Magaudda}, {Stelzer}, {Covey}, {Raetz},
  {Matt}, \& {Scholz}}]{Magaudda2020}
{Magaudda}, E., {Stelzer}, B., {Covey}, K.~R., {et~al.} 2020, \aap, 638, A20

\bibitem[{{Matsumoto}(2021)}]{Matsumoto2021}
{Matsumoto}, T. 2021, \mnras, 500, 4779

\bibitem[{{McIntosh} {et~al.}(2011){McIntosh}, {de Pontieu}, {Carlsson},
  {Hansteen}, {Boerner}, \& {Goossens}}]{2011Natur.475..477M}
{McIntosh}, S.~W., {de Pontieu}, B., {Carlsson}, M., {et~al.} 2011, \nat, 475,
  477

\bibitem[{{Meyer}(1985)}]{Meyer}
{Meyer}, J.~P. 1985, \apjs, 57, 173

\bibitem[{{M{\"u}ller} {et~al.}(2003){M{\"u}ller}, {Hansteen}, \&
  {Peter}}]{2003A&A...411..605M}
{M{\"u}ller}, D.~A.~N., {Hansteen}, V.~H., \& {Peter}, H. 2003, \aap, 411, 605

\bibitem[{Murphy(1985)}]{Murphy}
Murphy, R.~J. 1985, PhD thesis, University of Maryland

\bibitem[{{Parker}(1972)}]{parker1972}
{Parker}, E.~N. 1972, \apj, 174, 499

\bibitem[{{Parker}(1983)}]{parker1983}
{Parker}, E.~N. 1983, \apj, 264, 642

\bibitem[{{Pencil Code Collaboration} {et~al.}(2021){Pencil Code
  Collaboration}, {Brandenburg}, {Johansen}, {Bourdin}, {Dobler}, {Lyra},
  {Rheinhardt}, {Bingert}, {Haugen}, {Mee}, {Gent}, {Babkovskaia}, {Yang},
  {Heinemann}, {Dintrans}, {Mitra}, {Candelaresi}, {Warnecke},
  {K{\"a}pyl{\"a}}, {Schreiber}, {Chatterjee}, {K{\"a}pyl{\"a}}, {Li},
  {Kr{\"u}ger}, {Aarnes}, {Sarson}, {Oishi}, {Schober}, {Plasson}, {Sandin},
  {Karchniwy}, {Rodrigues}, {Hubbard}, {Guerrero}, {Snodin}, {Losada},
  {Pekkil{\"a}}, \& {Qian}}]{PC2020}
{Pencil Code Collaboration}, {Brandenburg}, A., {Johansen}, A., {et~al.} 2021,
  The Journal of Open Source Software, 6, 2807

\bibitem[{{Pesnell} {et~al.}(2012){Pesnell}, {Thompson}, \&
  {Chamberlin}}]{2012SoPh..275....3P}
{Pesnell}, W.~D., {Thompson}, B.~J., \& {Chamberlin}, P.~C. 2012, \solphys,
  275, 3

\bibitem[{{Peter} {et~al.}(2004){Peter}, {Gudiksen}, \& {Nordlund}}]{peter2004}
{Peter}, H., {Gudiksen}, B.~V., \& {Nordlund}, {\AA}. 2004, \apjl, 617, L85

\bibitem[{{Pevtsov} {et~al.}(2003){Pevtsov}, {Fisher}, {Acton}, {Longcope},
  {Johns-Krull}, {Kankelborg}, \& {Metcalf}}]{Pevtsov}
{Pevtsov}, A.~A., {Fisher}, G.~H., {Acton}, L.~W., {et~al.} 2003, \apj, 598,
  1387

\bibitem[{{Pizzolato} {et~al.}(2003){Pizzolato}, {Maggio}, {Micela},
  {Sciortino}, \& {Ventura}}]{Pizzolato2003}
{Pizzolato}, N., {Maggio}, A., {Micela}, G., {Sciortino}, S., \& {Ventura}, P.
  2003, \aap, 397, 147

\bibitem[{{Rappazzo} {et~al.}(2008){Rappazzo}, {Velli}, {Einaudi}, \&
  {Dahlburg}}]{Rappazzo2008}
{Rappazzo}, A.~F., {Velli}, M., {Einaudi}, G., \& {Dahlburg}, R.~B. 2008, \apj,
  677, 1348

\bibitem[{{Reiners}(2012)}]{Reiners}
{Reiners}, A. 2012, Living Reviews in Solar Physics, 9, 1

\bibitem[{{Reiners} {et~al.}(2014){Reiners}, {Sch{\"u}ssler}, \&
  {Passegger}}]{Reiners2014}
{Reiners}, A., {Sch{\"u}ssler}, M., \& {Passegger}, V.~M. 2014, \apj, 794, 144

\bibitem[{{Rempel}(2017)}]{Rempel17}
{Rempel}, M. 2017, \apj, 834, 10

\bibitem[{{Rosner} {et~al.}(1978){Rosner}, {Tucker}, \& {Vaiana}}]{Rosner}
{Rosner}, R., {Tucker}, W.~H., \& {Vaiana}, G.~S. 1978, \apj, 220, 643

\bibitem[{{Saar} \& {Brandenburg}(2001)}]{Saar2001}
{Saar}, S. \& {Brandenburg}, A. 2001, in Astronomical Society of the Pacific
  Conference Series, Vol. 248, Magnetic Fields Across the Hertzsprung-Russell
  Diagram, ed. G.~{Mathys}, S.~K. {Solanki}, \& D.~T. {Wickramasinghe}, 231

\bibitem[{{Saar}(1996)}]{Saar1996}
{Saar}, S.~H. 1996, in Stellar Surface Structure, ed. K.~G. {Strassmeier} \&
  J.~L. {Linsky}, Vol. 176, 237

\bibitem[{{Schou} {et~al.}(2012){Schou}, {Scherrer}, {Bush}, {Wachter},
  {Couvidat}, {Rabello-Soares}, {Bogart}, {Hoeksema}, {Liu}, {Duvall}, {Akin},
  {Allard}, {Miles}, {Rairden}, {Shine}, {Tarbell}, {Title}, {Wolfson},
  {Elmore}, {Norton}, \& {Tomczyk}}]{HMI}
{Schou}, J., {Scherrer}, P.~H., {Bush}, R.~I., {et~al.} 2012, \solphys, 275,
  229

\bibitem[{{Schrijver}(1983)}]{1983A&A...127..289S}
{Schrijver}, C.~J. 1983, \aap, 127, 289

\bibitem[{{Schrijver}(1987)}]{Schrijver1987}
{Schrijver}, C.~J. 1987, \aap, 172, 111

\bibitem[{{Spitzer}(1962)}]{Spitzer}
{Spitzer}, L. 1962, {Physics of Fully Ionized Gases} (2nd edition (New York:
  Interscience))

\bibitem[{{Testa} {et~al.}(2015){Testa}, {Saar}, \&
  {Drake}}]{2015RSPTA.37340259T}
{Testa}, P., {Saar}, S.~H., \& {Drake}, J.~J. 2015, Philosophical Transactions
  of the Royal Society of London Series A, 373, 20140259

\bibitem[{{Tomczyk} {et~al.}(2007){Tomczyk}, {McIntosh}, {Keil}, {Judge},
  {Schad}, {Seeley}, \& {Edmondson}}]{Tomczyk}
{Tomczyk}, S., {McIntosh}, S.~W., {Keil}, S.~L., {et~al.} 2007, Science, 317,
  1192

\bibitem[{{van Ballegooijen} {et~al.}(2011){van Ballegooijen}, {Asgari-Targhi},
  {Cranmer}, \& {DeLuca}}]{Balle2011}
{van Ballegooijen}, A.~A., {Asgari-Targhi}, M., {Cranmer}, S.~R., \& {DeLuca},
  E.~E. 2011, \apj, 736, 3

\bibitem[{{Vernazza} {et~al.}(1981){Vernazza}, {Avrett}, \&
  {Loeser}}]{Vernazza1981}
{Vernazza}, J.~E., {Avrett}, E.~H., \& {Loeser}, R. 1981, \apjs, 45, 635

\bibitem[{{Vidotto} {et~al.}(2014){Vidotto}, {Gregory}, {Jardine}, {Donati},
  {Petit}, {Morin}, {Folsom}, {Bouvier}, {Cameron}, \& {Hussain}}]{Vidotto}
{Vidotto}, A.~A., {Gregory}, S.~G., {Jardine}, M., {et~al.} 2014, \mnras, 441,
  2361

\bibitem[{{Warnecke} \& {Bingert}(2020)}]{Warnecke}
{Warnecke}, J. \& {Bingert}, S. 2020, Geophysical and Astrophysical Fluid
  Dynamics, 114, 261

\bibitem[{{Warnecke} \& {Peter}(2019{\natexlab{a}})}]{Warnecke2019}
{Warnecke}, J. \& {Peter}, H. 2019{\natexlab{a}}, A\&A, 624, L12

\bibitem[{{Warnecke} \& {Peter}(2019{\natexlab{b}})}]{Warnecke2020}
{Warnecke}, J. \& {Peter}, H. 2019{\natexlab{b}}, arXiv e-prints,
  arXiv:1910.06896

\bibitem[{{Withbroe} \& {Noyes}(1977)}]{withbroe}
{Withbroe}, G.~L. \& {Noyes}, R.~W. 1977, \araa, 15, 363

\bibitem[{{Wright} {et~al.}(2011){Wright}, {Drake}, {Mamajek}, \&
  {Henry}}]{Wright2011}
{Wright}, N.~J., {Drake}, J.~J., {Mamajek}, E.~E., \& {Henry}, G.~W. 2011,
  \apj, 743, 48

\bibitem[{{Zhuleku} {et~al.}(2020){Zhuleku}, {Warnecke}, \& {Peter}}]{Zhuleku}
{Zhuleku}, J., {Warnecke}, J., \& {Peter}, H. 2020, \aap, 640, A119

\end{thebibliography}

\Online
\appendix

\section{Quenching of convective motions}
\label{A:quen}

The current numerical experiments presented in \sect{S.exp} might overestimate the amount of Poynting flux injected from the bottom boundary.
This is because, at the surface, (strong) magnetic fields would quench the convective motions.
To properly account for this effect, the near-surface magneto-convection would have to be included self-consistently into the model.
Still, we can estimate what effect this quenching would have on the power-law indices we derive.

The horizontal velocities at the surface, $u_{\rm h}$, that drive the magnetic field will be reduced in the strong-field regions according to
\begin{equation}
     \label{E:uh.quenched}
      u_{\rm h} \mapsto u_{\rm h}^\prime=q\,u_{\rm h}.
\end{equation}
The quenching function $q$ can be expressed as a function of the vertical magnetic field at the surface, $B_{z,0}$,
\begin{equation}
    \label{E:quenching}
    q=f(|B_{z,0}|)
    \qquad \mbox{with} \qquad
    \left\{
    \begin{array}{ll}
    q=1:             &  |B_{z,0}| \ll B_{\rm{eq}} ,\\
    q=q_{\rm{sat}}:  &  |B_{z,0}| \gg B_{\rm{eq}} .
    \end{array}
    \right.
\end{equation}
Essentially, if the magnetic field is well below the equipartition field strength, $B_{\rm{eq}}$, there is no quenching, while for large $|B_{z,0}|$ the surface velocities are reduced by the factor $q_{\rm{sat}}$.
For our experiments, we assume values of $q_{\rm{sat}}$ of 1, 2/3, and 1/3.

For a smooth transition of the quenching $q$ from $1$ to $q_{\rm{sat}}$ we apply a cubic step function that depends on the logarithm of the magnetic field. We assume that the inflection point of the function $q$ is at the equipartition field strength, $B_{\rm{eq}}$. In our models this is (on average) 2.3\,kG. The transition from $q{=}1$ to $q_{\rm{sat}}$ happens over a range of 0.8 in $\log_{10}|B_{z,0}|$. 
This quenching as a function of magnetic field (for the case $q_{\rm{sat}}{=}1/3$) is very close to the quenching used by \cite{GN05b} as defined in their Eq.\,12.

\begin{figure}
\includegraphics[width=\columnwidth]{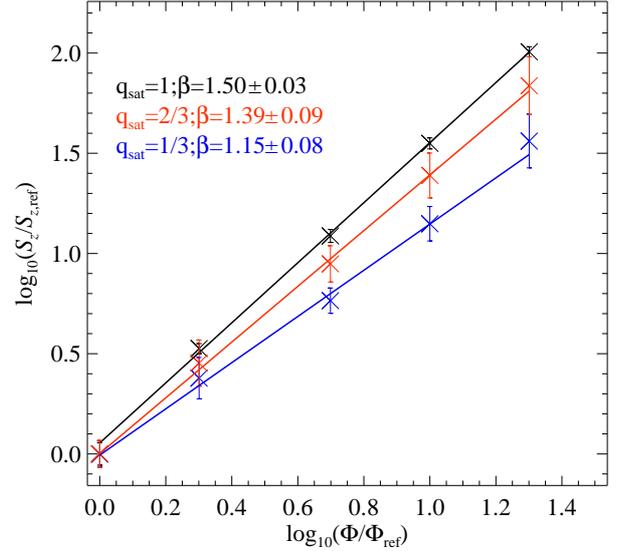}
\caption{Effect of quenching of horizontal motions by strong magnetic field.
We show the relation of the quenched Poynting flux $\widetilde{S}_z$ as defined in \eqn{E:Sz.hor}  to the unsigned surface magnetic flux $\Phi$. Each data point represents an average in time and space for each run. The values are normalized to model run 1B represented here by the subscript 'ref'. The black line shows the power-law fit to the case where no quenching is included (i.e. $q_{\rm{sat}}=1$). The red and blue lines show power-law fits for $q_{\rm{sat}}=2/3$ and 1/3. The corresponding power-law indices $\beta$ for $\widetilde{S}_z\propto\Phi^{\,\beta}$ are given with the legend.
}
\label{fig:A}
\end{figure}

In general, the Poynting flux is defined as
\begin{equation}
S_{z}=\eta(\boldsymbol{j}\times \boldsymbol{B})\big |_{z}-\frac{1}{\mu_{0}}(\boldsymbol{u}\times\boldsymbol{B}\times
\boldsymbol{B})\big |_{z}.
\label{A.pz}
\end{equation}
The first term $\boldsymbol{j}\times \boldsymbol{B}$ is not affected by the quenching of horizontal motions. 
The second term can be further decomposed into two terms,
\begin{equation}
    \boldsymbol{u}\times \boldsymbol{B}\times \boldsymbol{B}\big |_{z}~=~u_z\, B^2_{\rm h}~-~B_z\, B_{\rm h}\, u'_{\rm h} ,
    \label{A:p}
\end{equation}
where we now use the horizontal velocities quenched by the magnetic field, $u'_{\rm h}$.
The subscript h refers to horizontal component combining the $x$ and $y$ direction.
From the two terms in \eqn{A:p} the first term is less relevant in our context, because the vertical velocities are smaller than the horizontal ones at the surface, and because the horizontal magnetic field is smaller than the vertical one. 
Only the second term,
\begin{equation}
\label{E:Sz.hor}
\widetilde{S}_z= \frac{1}{\mu_{0}}~B_z\, B_{\rm h}\, u'_{\rm h},
\end{equation}
will be affected by quenching of the horizontal velocities and is of foremost interest in the discussion in this Appendix.

To test the effect of the quenching of the horizontal motions on the Poynting flux we therefore evaluate $\widetilde{S}_z$ from \eqn{E:Sz.hor}. To this end, we calculate the quenching term $q$ as defined in \eqn{E:quenching} at each location at the bottom boundary, i.e., the surface, calculate the quenched velocity according to \eqn{E:uh.quenched}, and then derive the vertical component of the Poynting flux due to the quenched horizontal motions, $\widetilde{S}_z$, at the bottom boundary of our model runs. 
As done for the other (average) quantities in our study, the quenched Poynting flux at the bottom is (horizontally) averaged in space and in time during the relaxed state of the simulations.
For better comparison between the different models, we normalize the (average) magnetic flux and Poynting flux at the bottom by the respective values from model run 1B.

To quantify the effect of the quenching on the scaling between magnetic flux $\Phi$ and Poynting flux at the surface $\widetilde{S}_z$, we conduct experiments for three cases with $q_{\rm{sat}}=1$, 2/3, and 1/3.
These experiments are illustrated in \fig{fig:A}.
For the case with no quenching ($q_{\rm{sat}}{=}1$), we find a power-law scaling  $\widetilde{S}_z\propto\Phi^{\,\beta}$ with a power-law index $\beta=1.5$.
This is slightly smaller than $\beta$ in the full model in \fig{Figscaling} and \eqn{E:num.S.vs.Phi}, but still within the range of uncertainties given there. 
The small difference is because of neglecting the other terms of the Poynting flux in \eqn{A.pz} and because in the main text we consider the Poynting flux at the base of the corona (see \sect{S.heat}) and not at the bottom boundary as done here.
More importantly, the power-law relation $\widetilde{S}_z\propto\Phi^{\,\beta}$ gets more shallow for stronger quenching, with the power-law index $\beta$ going down to about 1.4 and 1.15 for values of $q_{\rm{sat}}$ of 2/3 and 1/3.
Consequently, for a more realistic setup with a quenching similar to \cite{GN05b}, i.e., $q_{\rm{sat}}=1/3$,  we would find a reduction of the power-law index $\beta$ by about 25\% compared to the case of no quenching.

The reduction of the power-law index $\beta$ by the quenching of horizontal motions will also affect how the X-ray luminosity $L_{\rm{X}}$ scales with the surface magnetic flux $\Phi$.
In the framework of the analytical model in \sect{Lx.vs.Phi.ana}, the power-law index $m$ of $L_{\rm{X}}\propto \Phi^m$ is expressed in \eqn{E.ms} as a function of three parameters, $\alpha, \beta, \gamma$, of which only the parameter $\beta$ will be affected by the quenching. $\alpha$ and $\gamma$  will retain the values discussed in \sect{Lx.vs.Phi.ana}.
Following \eqn{E.ms}, $m$ is linear in $\beta$. Hence its reduction through the quenching for a more realistic setup would be the same as for $\beta$, that is, about 25\%.
Therefore we would expect the quenching to reduce $m$ as found in our numerical experiments as displayed in \fig{Figscalineuv} and discussed in \sect{Lx.vs.Phi} to about $m=2.6$.
This would be very close to the value of 2.68 found by \cite{Kochukhov} in their recent observational study, but still considerably higher than the traditional value that is assumed to be not considerably higher than unity \citep{Fisher,Pevtsov}.

\end{document}